\documentclass[twocolumn,aps,showpacs,prb,tightenlines,amsmath,amssymb]{revtex4}
\usepackage{graphicx}    
\usepackage{amssymb}  
\usepackage{dcolumn}
\usepackage{amsmath}     
\usepackage{bm}                                          
\usepackage{colordvi}                   
\newcommand{\bgreek}[1]{\mbox{\boldmath$#1$\unboldmath}}
\begin{document}             
\title{Energy spectra of three electrons in Si/SiGe single and vertically
  coupled double quantum dots}
\author{Z. Liu}
\affiliation{Hefei National Laboratory for Physical Sciences at Microscale and
  Department of Physics, University of Science and Technology of China, Hefei,
  Anhui, 230026, China} 
\author{L. Wang}     
\affiliation{Hefei National Laboratory for Physical Sciences at Microscale and
  Department of Physics, University of Science and Technology of China, Hefei,
  Anhui, 230026, China} 
\author{K. Shen}
\thanks{Author to  whom correspondence should be addressed}
\email{kkshen@mail.ustc.edu.cn.}
\affiliation{Hefei National Laboratory for Physical Sciences at Microscale and
  Department of Physics, University of Science and Technology of China, Hefei,
  Anhui, 230026, China} 
\date{\today}

\begin{abstract}
 We study three-electron energy spectra in Si/SiGe single and
 vertically coupled double quantum dots where all the relevant
 effects, such as, the Zeeman
 splitting, spin-orbit coupling, valley coupling and electron-electron Coulomb
 interaction are explicitly
 included.  In the absence of magnetic field, our results in single quantum dots
 agree well with the experiment by Borselli {\em et al.}
 [Appl. Phys. Lett. {\bf 98}, 123118 (2011)]. We identify the spin and valley
 configurations of the ground state in the experimental cases and give a complete
 phase-diagram-like picture of the ground state configuration with respect to the dot
 size and valley splitting. We also explicitly investigate the three-electron
 energy spectra of the pure and mixed valley configurations with magnetic fields in
 both Faraday and Voigt configurations. We find that the ground state
 can be switched between doublet and quartet by tuning the
 magnetic field and/or dot size.
 The three-electron energy spectra present
 many anticrossing points between different spin states due to the spin-orbit
 coupling, which are expected to
 benefit the spin manipulation. We show that the
 negligibly small intervalley Coulomb
 interaction can result in magnetic-field independent quartet-doublet
 degeneracy in the three-electron energy spectrum of the mixed valley
 configuration. Furthermore, we study the barrier-width and barrier-height
 dependences in vertically coupled double quantum
 dots with both pure and mixed valley configurations. Similar to the single
 quantum dot case, anticrossing behavior and quartet-doublet
 degeneracy are observed. 
\end{abstract}


\maketitle
\section{INTRODUCTION}

Silicon quantum dots (QDs) are proposed to be prominent candidates for spin
qubits owing to the long spin-decoherence time,\cite{Loss1,Reimann1,Hanson2,Wu_rev,Eriksson1,Berezovsky1,Harju1,Dimi1,Dimi2,Lwang1,Lwang2,Dimi3,Tahan1}
which is of great importance for coherent manipulation, information storage and
quantum error correction.\cite{Tahan1,Shor1,Sherwin1,Cory1,Bennett1}
The long decoherence time in Si-based devices results from the weak hyperfine
interaction,\cite{Lwang1} small spin-orbit coupling
(SOC)\cite{Dresselhaus1,Vervoort1,Vervoort2,Rashba1,Bockelmann1,Landolt1,Nestoklon1} and
weak electron-phonon interaction.\cite{Li1} As an indirect-gap semiconductor, the
conduction band of the bulk Si has six degenerate minima or valleys. This
six-fold degeneracy can be lifted by strain or quantum confinement, e.g.,
it separates into a four-fold degeneracy and a
two-fold one in [001] quantum wells. The two-fold degenerate valleys with low
energy can be further lifted by a valley splitting due to the interface
scattering.\cite{Friesen1,Ando1} The existence of the valley degree of freedom
makes the Si-based qubits more attractive.\cite{Dimi1,Dimi2,Lwang1,Lwang2,Dimi3}
Moreover, the mature microfabrication
technology of the classical Si-based electronics
is proposed to benefit the realization of Si spin qubits.\cite{Dimi1,Lwang1}

Recently, Si quantum dots (QDs) have been
widely investigated both experimentally 
and theoretically.\cite{Harju1,Eriksson1,Borselli1,Xiao1,Lim1,Lim2,Fabian1,
  Dimi1,Dimi2,Dimi3,Lwang1,Lwang2,Barnes1}
In the experiments, Si metal-on-semiconductor and Si/SiGe QDs with a
tunable electron filling number from zero have been
fabricated, where the valley splitting, few-electron
energy spectrum and spin relaxation time have been measured.\cite{Borselli1,Xiao1,Lim1,Lim2}
The theoretical works mainly focus on the one-electron Zeeman
sublevels or the singlet-triplet states of two
electrons.\cite{Harju1,Eriksson1,Fabian1,Dimi1,Dimi2,Dimi3,Lwang1,Lwang2}
For example, Culcer {\em et al.}\cite{Dimi1,Dimi2,Dimi3} analyzed the
initialization and manipulation of one-electron and two-electron qubits in
lateral coupled double quantum dots (DQDs) by utilizing the valley degree of
freedom. Raith {\em et 
  al.}\cite{Fabian1} calculated the energy spectrum and the spin relaxation time
in one-electron QDs. By explicitly including the electron-electron Coulomb
interaction,  Wang {\em et al.}\@ obtained the two-electron spectrum from
exact-diagonalization method and calculated the
singlet-triplet relaxation time in both single\cite{Lwang1} and
lateral coupled double\cite{Lwang2} QDs. 
As pointed out by Barnes {\em et al.},\cite{Barnes1} the robustness of the
quantum states against charge impurity and noise in QDs can be improved by
increasing the number of the electrons, which reveals the necessity
of the theoretical investigation on multi-electron
spin qubits. The energy spectra in
multi-electron GaAs QDs have been
explicitly calculated to identify the specific spin configuration of each state,
where only one valley is relevant.\cite{GaAs1,GaAs2,GaAs3,GaAs4,Gamucci1}
However, to the best of our knowledge, there is no report on the explicit calculation
in silicon QDs with three or more electrons due to the complication of the
calculation. Alternatively, Hada {\em et al.}\cite{Hada1} neglected the
correlation effect and calculated the three-electron energy spectrum within
single configuration approximation in single QDs. However, the correlation
effect has been shown to present significant influence on the energy
spectrum via strong Coulomb interaction in Si QDs.\cite{Lwang1,Hada1,Eto1}
Therefore, in order to obtain the accurate convergent spectrum, the
exact-diagonalization method with sufficient basis functions is
required. The goal of the present work is to analyze the energy spectrum 
in the three-electron Si/SiGe QDs based on the exact-diagonalization method.

In this work, we calculate the three-electron energy spectrum in both single and
vertically coupled double QDs with the Zeeman splitting, SOC, valley
coupling and electron-electron Coulomb
interaction explicitly included. 
In the single dot case, we first calculate the ground state energy in the
absence of the magnetic field, where good agreement with the experimental data
is achieved. Our calculation also uncovers the valley and spin
configurations of the ground state in experiment. We present a complete
phase-diagram-like picture to describe the spin and valley configurations of the
ground state. We find that the ground state is of pure (mixed) valley
  configuration at large (small) valley splitting. The
spin configuration of 
the ground state can be controlled through dot size for the pure valley
configuration, while that in the mixed valley configuration is always
doublet. The magnetic field dependence of the
three-electron energy spectrum in each ``phase'' is investigated. 
We take into account both
orbital effect and Zeeman splitting for the perpendicular magnetic field
and only Zeeman splitting
for the parallel magnetic field owing to the strong
confinement along the growth direction. We find that the spin configuration of
the ground state can also be switched by
magnetic field. For the mixed valley
configuration, we find interesting quartet-doublet degeneracy, which results from
the negligibly small intervalley interaction. In the DQD case, the barrier-width
and barrier-height dependences of the energy spectrum with both pure and mixed valley
configurations are discussed.
Moreover, we show many anticrossing points
resulting from the SOCs in all cases.

This paper is organized as follows. In Sec.\,II, we set up our model
and formalism. In Sec.\,III,  we show our results of the three-electron energy
spectrum in single QDs and vertically coupled DQDs from the exact-diagonalization method.
We investigate the perpendicular and parallel magnetic-field dependences in single
QDs and the barrier-width and barrier-height dependences in
vertically coupled DQDs. The comparison with experiment in single QD case is
also given in this section. Finally, we summarize in Sec.\,IV. 

\section{MODEL AND FORMALISM}
We set up our model in a double quantum well along
[001]-direction. The confinement along this direction is described by\cite{YYWang1,Shen1}
\begin{eqnarray} V_{z}(z)&=& \left\{
    \begin{array}{cc}
      V_0 ,&\mbox{$|z|< \tfrac{a}{2}$},\\[3pt]
      0,&\mbox{$\tfrac{a}{2}\leq  |z|\leq \tfrac{a}{2}+d$},\\[3pt]
      \infty ,& \mbox{$\rm  otherwise$},
    \end{array} \right.
  \label{eq1}
\end{eqnarray}
with the inter-well barrier height denoting as $V_0$.
Here, $d$ and $a$ represent the width of each well and that of the
barrier, respectively. The lateral confinement is chosen to be
a parabolic potential $V_{\rm c}(x,y)=\tfrac{1}{2}m_t\omega_0^2(x^2+y^2)$
with $m_t$ representing the in-plane effective mass and $\omega_0$
being the confining potential frequency.\cite{Fock1,Darwin1}
The effective diameter $d_0$ is given by
$\sqrt{\hbar\pi/(m_t\omega_0)}$. The total confinement potential then can be written
as $V(\mathbf r)=V_{\rm c}(x,y)+V_{z}(z)$. For an infinitesimal barrier width
($a\sim 0$), our model reduces to the single dot case.

The external magnetic field with perpendicular and parallel components is
expressed by ${\mathbf
  B}=B_{\perp}\hat{\mathbf z}+B_{\parallel}\hat{\mathbf x}$.
The single-electron Hamiltonian reads\cite{Lwang1}
\begin{eqnarray} 
  H_{\rm e}  = \frac{P_x^2+P_y^2}{2m_t}+\frac{P_z^2}{2m_z}+V(\mathbf r)+H_{\rm
    Z}+H_{\rm so}+H_{\rm v},
  \label{eq2}
\end{eqnarray}
where $m_z$ denotes the effective mass along the growth direction and ${\mathbf
  P}=-i\hbar{\bgreek \nabla}+(e/c){\mathbf A}$ with the vector potential ${\mathbf
  A}=(-yB_{\perp},xB_{\perp},2yB_{\parallel})/2$. In our calculation, the
orbital effect of the parallel component of the magnetic field is neglected due
to the strong confinement along the $z$-direction
and the vector potential in the mechanical momentum then reduces to
${\bf A}=(-yB_{\perp},xB_{\perp},0)/2$. In Eq.\,(\ref{eq2}), the Zeeman 
splitting is given by $H_{\rm
  Z}=\tfrac{1}{2}g\mu_B(B_{\perp}\sigma_z+B_{\parallel}\sigma_x)$ with
$g$ being the Land$\acute{\rm e}$ factor. The SOCs, including the Rashba
term\cite{Rashba1} due to the structure
inversion asymmetry (SIA) and the interface-inversion asymmetry (IIA)
term,\cite{Vervoort2,Vervoort1,Nestoklon1} are expressed by
\begin{eqnarray} 
  H_{\rm so}  = a_0(P_x\sigma_y-P_y\sigma_x)+b_0(-P_x\sigma_x+P_y\sigma_y),\mbox{}
  \label{eq3} 
\end{eqnarray}
where  $a_0$ ($b_0$) represents the coupling coefficient of the Rashba (IIA)
term. $H_{\rm v}$ in Eq.~(\ref{eq2}) describes the 
coupling between the two low-energy valleys lying at $\pm\langle k_{\rm Si}
\rangle$ along the $z$-axis with $\langle k_{\rm Si}
\rangle=0.85(2\pi/a_{\rm Si})$.\cite{Friesen1} Here, $a_{\rm Si}=5.43\;{\rm \AA}$
stands for the lattice constant of silicon. In this work, we use ``$z$''
(``$\overline{z}$'') to denote the valley lying at 
$\langle k_{\rm Si} \rangle$ ($-\langle k_{\rm Si} \rangle$) for convenience.

In order to build up a complete set of single-electron basis functions,
we define $H'_0=H_0+H_{\rm v}$ with
$H_0=\tfrac{P_x^2+P_y^2}{2m_t}+\tfrac{P_z^2}{2m_z}+V(\mathbf
r)$. Due to the separation of the lateral and vertical confinement, one can
easily solve the Schr\"odinger equation of $H_0$.
From the lateral part, one obtains the eigenvalues\cite{Cheng1,Fock1,Darwin1}
\begin{eqnarray}
  E_{nl}=\hbar\Omega(2n+|l|+1)+\hbar l\omega_B,\mbox{}
 \label{eq4}
\end{eqnarray}
where $\Omega=\sqrt{\omega_0^2+\omega_B^2}$ and
$\omega_B=eB_{\perp}/(2m_t)$. The eigenfunctions read\cite{Cheng1,Fock1,Darwin1}
\begin{eqnarray}
  K_{nl}(r,\theta)=N_{n,l}(\alpha r)^{|l|}e^{-(\alpha
    r)^2/2}e^{il\theta}L_n^{|l|}[(\alpha  r)^2],\mbox{}
  \label{eq5}
\end{eqnarray}
with $N_{n,l}=\{\alpha^2n!/[\pi (n+|l|)!]\}^{1/2}$ and
$\alpha=\sqrt{m_t\Omega/\hbar}$. $L^{|l|}_{n}$ is the generalized Laguerre
polynomial. Here, $n=0,1,2,...$ is the radial quantum number and
$l=0,\pm1,\pm2,...$ is the azimuthal angular momentum quantum number.
For the vertical part, we include the lowest two subbands, one with even parity
(denoted by subscript $n_z=0$) and the other with odd parity
($n_z=1$). The corresponding wave functions can be expressed as\cite{Shen1}
\begin{eqnarray} \xi_0(z)&=& \left\{
    \begin{array}{ccccccc}
      \nonumber C_0\sin[k_0(z-\tfrac{a}{2}-d)] ,&\mbox{$\tfrac{a}{2}\le z
        \leq\tfrac{a}{2}+d$} , \\[3pt] 
      \nonumber     A_0\cosh(\beta_0 z),&\mbox{$|z|< \frac{a}{2}$},\\[3pt]
      \nonumber     C_0\sin[k_0(-z-\tfrac{a}{2}-d)] ,&\mbox{$-\tfrac{a}{2}-d\le z \leq
        -\tfrac{a}{2}$},\\[3pt]
      0, & \mbox{otherwise},\\
    \end{array} \right.\\
  \label{eq6}
\end{eqnarray}
and
\begin{eqnarray} \xi_1(z)&=& \left\{
    \begin{array}{ccccccc}
      \nonumber     C_1\sin[k_1(z-\tfrac{a}{2}-d)] ,&\mbox{$\tfrac{a}{2}\le z
        \leq\tfrac{a}{2}+d$} ,\\[3pt] 
\nonumber     A_1\sinh(\beta_1 z),&\mbox{$|z|< \tfrac{a}{2}$},\\[3pt]
\nonumber     C_1\sin[k_1(z+\tfrac{a}{2}+d)] ,&\mbox{$-\tfrac{a}{2}-d\le z \leq
       -\tfrac{a}{2}$},\\[3pt]
0, &\mbox{otherwise},\\
   \end{array} \right.\\
  \label{eq7}
\end{eqnarray}
with $k_{n_z}=\sqrt{2m_zE_{n_z}/\hbar^2}$ and
$\beta_{n_z}=\sqrt{2m_z(V_0-E_{n_z})/\hbar^2}$. $E_{n_z}$ is the
eigenvalue of the $n_z$-th subband. For the single dot case, only the
lowest subband ($n_z=0$) is relevant. With the knowledge of the eigenfunctions
of $H_0$, one expresses the single-electron basis functions in different
valleys as $\phi^{z(\overline{z})}_{nln_z}=K_{nl}\xi_{n_z}e^{\pm 
  ik_{\rm Si}z}u_{z(\overline{z})}({\mathbf r})$, with $u_{z(\overline{z})}(\mathbf r)$
being the lattice-periodic Bloch functions.

Then we introduce the valley coupling $H_{\rm v}$ according to
Ref.~\onlinecite{Friesen1}, where the relevant components are given
by\cite{Dimi1,Lwang1}
$\langle\phi^{z}_{n^\prime l^\prime n_z}|H_{\rm
  v}|\phi^{z}_{nln_z}\rangle= \langle\phi^{\overline{z}}_{n^\prime l^\prime n_z}|H_{\rm
  v}|\phi^{\overline{z}}_{nln_z}\rangle =\Delta^0_{n_z,n_z}\delta_{n,n'}\delta_{l,l'}$
and
$\langle\phi^{z}_{n^\prime l^\prime n_z'}|H_{\rm
  v}|\phi^{\overline{z}}_{nln_z}\rangle=\langle\phi^{\overline{z}}_{n^\prime
  l^\prime n_z'}|H_{\rm
  v}|\phi^{z}_{nln_z}\rangle=\Delta^1_{n_z,n_z'}\delta_{n,n'}\delta_{l,l'}$.
The expressions of $\Delta^0_{n_z,n_z}$ and $\Delta^1_{n_z,n_z'}$ are
given in Appendix~\ref{appA}\@. Since the valley coupling between
different subbands  ($n_z\neq n_z^\prime$) in our case is much smaller than
the intersubband energy difference ($<2$~\%), we neglect the
intersubband valley coupling. The eigenstates of $H_0^\prime$ then can be written as
$\phi^{\pm}_{nln_z}=\tfrac{1}{\sqrt{2}}(\phi^z_{nln_z}\pm\phi^{\overline{z}}_{nln_z})$
and the corresponding eigen-energies are
$E^{\pm}_{nln_z}=E_{nl}+E_{n_z}+E^{\pm}_{n_z}$.
with $E^{\pm}_{n_z}=\Delta^0_{n_z,n_z}\pm
|\Delta^1_{n_z,n_z}|$ representing the energy from the valley degree
of freedom and $\Delta E^{\rm v}_{n_z}=2|\Delta^1_{n_z,n_z}|$ being the valley splitting.

For the three-electron case, the total Hamiltonian can be expressed as
\begin{eqnarray} 
  H_{\rm tot} & = & H^1_{\rm e} + H^2_{\rm e} + H^3_{\rm e} + H^{12}_{\rm C} +
  H^{23}_{\rm C} + H^{13}_{\rm C}.\mbox{}
  \label{eq8}
\end{eqnarray}
The superscripts ``1'', ``2'' and ``3'' on the right hand side of the
equation label the relevant electrons, e.g., $H^i_{\rm e}$
represents the single-electron Hamiltonian of the $i$-th electron given by
Eq.~(\ref{eq2}) and $H^{i,j}_C=\tfrac{e^2}{4\pi\varepsilon_0\kappa|{\mathbf
    r_i}-{\mathbf r_j}|}$
stands for the Coulomb interaction between $i$-th and $j$-th electrons.
Here, $\kappa$ is the relative static dielectric constant.

By using the single-electron functions \{$\phi^{\pm}_{nln_z}$\} or 
$\{|nln_{z}n_{{\rm v}} \rangle\} $ ($n_{\rm v}=\pm$
denotes the valley eigenfunction),
we construct the three-electron basis functions in the form of either doublet
($S_{\rm tot}=\frac{1}{2}$, denoted 
as $|D^{(\Xi)}_{S^*}\rangle$) or quartet ($S_{\rm tot}=\frac{3}{2}$, denoted as
$|Q^{(\Xi)}_{S^*}\rangle$) with Clebsch-Gordan
coefficients. Subscript $S^*$ stands for the spin magnetic
quantum number. Superscript $\Xi$ denotes four valley configurations 
of three-electron basis functions, i.e. $\Xi=-(+)$ for the case three
electrons in ``$-$'' (``$+$'') valley and $\Xi={\rm m}$ ($\widetilde{\rm m}$) for two
electrons in ``$-$''(``$+$'') valley and one electron in ``$+$''(``$-$'') valley.
The details of the three-electron basis functions are
given in Appendix~\ref{appB}\@.

One calculates the matrix elements of Eq.\,(\ref{eq8}) under the three-electron basis
functions. The details of the Coulomb interaction are given in
Appendix\,\ref{appC}. By neglecting the small coupling between basis
functions of different valley configurations,\cite{Dimi1,Lwang1}
the complete basis functions can be divided into 
four individual sets according to valley configurations ($\Xi=-$,
${\rm m}$, $\widetilde{\rm m}$ and $+$). We diagonalize the Hamiltonian of each
subspace and define an eigenstate as doublet
$|D^{(\Xi)}_{S^*}\rangle$ (quartet $|Q^{(\Xi)}_{S^*}\rangle$) if its amplitude of
doublet (quartet) components is greater than $50$~$\%$. In the following, we
still use the notations
$|D^{(\Xi)}_{S}\rangle$ and
$|Q^{(\Xi)}_{S}\rangle$
to describe the spin properties of the eigenstates without any confusion.

\section{NUMERICAL RESULTS}

In our calculation, the effective mass $m_t=0.19 m_0$ and $m_z=0.98 m_0$ with
$m_0$ representing the free electron mass.\cite{Dexter1} The strengths of the
SOCs between the states with the same valley index ``$\pm$'' are
taken as $a_0=\pm 6.06$~m/s and $b_0=\pm 30.31$~m/s (Ref.\,\onlinecite{Nestoklon1}). The
Land\'e factor $g=2$ (Ref.\,\onlinecite{Graefi1}) and the relative static dielectric
constant $\kappa=11.9$ (Ref.\,\onlinecite{Hada1}).
In single QDs, we take 1430 quartets and 3330 doublets to guarantee
the convergence of the energy spectrum, while in vertically coupled DQDs we
take 7316 quartets and 16008 doublets correspondingly.

\subsection{Single quantum dots}
\subsubsection{Comparison with experiment}
Recently, Borselli {\em et al.} investigated  four
Si/SiGe single QDs, which are labeled as Si1, Si2, Si3 and Si4
separately, in the absence of the magnetic field.\cite{Borselli1} They measured
the tuning voltages 
for the injection of an additional electron into the QDs, i.e., $\Delta
V^{N_{\rm e}}_G=V_G^{N_{\rm e}\leftrightarrow(N_{\rm e}+1)}-V_G^{(N_{\rm
    e}-1)\leftrightarrow N_{\rm e}}$, where $V^{N_{\rm
    e}\leftrightarrow(N_{\rm e}+1)}_G$ represents the gate voltage
where the electron number changes between $N_{\rm e}$ and
$N_{\rm e}$+1. Then the addition energy of an incoming electron was determined
by $\Delta \mu_{N_{\rm e}}=\mu_{N_{\rm e}+1}-\mu_{N_{\rm
    e}}=\alpha\Delta V^{N_{\rm e}}_G$, where $\mu_{N_{\rm e}} = E^{N_{\rm
    e}}_{\rm Tot}-E^{N_{\rm  e}-1}_{\rm Tot}$ with $E^{N_{\rm e}}_{\rm Tot}$
and $\alpha$ being the total energy of the $N_{\rm e}$-electron ground state and the
energy-voltage conversion factor. The experimental data of $\Delta \mu_1$
and $\Delta \mu_2$ of the four samples (noted with the superscript $\ast$) are listed in
Table\,\ref{table1}. 
\begin{table}
\begin{tabular}{lllll}
  \hline
  \hline \\[-6pt]
  \quad\quad\quad\quad\quad\quad\quad\quad &Si1 \quad \quad\quad\quad&
   Si2  \quad\quad\quad\quad& Si3 \quad\quad\quad \quad& Si4 \quad\quad\\[3pt]
  \hline \\[-6pt]
  $\Delta E^{\rm v}_{0}$ (meV)        &0  &0  &0.27  &0.12  \\[3pt]
  $d$  (nm)       &3.945  &3.945  &3.945  &3.926   \\[3pt]
  $\Delta\mu^*_1$ (meV) \quad&4.520         &3.800  &3.916   &4.680  \\[3pt]
  $\Delta\mu_1 $  (meV)  \quad&4.452        &3.828 &4.008   &4.671 \\[3pt]
  $R_1$ (\%)              \quad&1.5    &0.7  & 2.3    & 0.2  \\[3pt]
  $\Delta\mu^*_2$ (meV)\quad&3.226  & 2.863 &3.146   &3.594 \\[3pt]
 $\Delta\mu_2$    (meV) &3.318   & 2.824  & 3.061  &3.613 \\[3pt]
 $R_2$ (\%)        &2.9    & 1.4  & 2.7  &0.5    \\[3pt]
 $d_0$ (nm)         &31.5    & 35.3  &34.1  &30.3    \\[3pt]
 $\Xi$         &${\rm m/\widetilde m}$   &${\rm m/\widetilde m}$  &$-$
                    &${\rm m}$\\[3pt] 
$S_{\rm tot}$         &$1/2$   &$1/2$  &$3/2$  &$1/2$\\[3pt]
Deg.   & 4& 4& 4& 2\\[3pt]
\hline
\hline
\end{tabular}
\caption{The comparison between experimental data\cite{Borselli1} and our
  results in four devices, Si1-Si4.
  $\Delta E^{\rm v}_{0}$ and $d$ are the valley splitting and
 half-well width used in the calculation.
 $\Delta\mu_i^*$ represent the experimental addition energy
 and $\Delta\mu_i$ stand for the theory results. $R_i$ stand for the
 relative error between  $\Delta\mu_i^*$ and $\Delta\mu_i$. The effective diameter $d_0$ is
 determined from  our calculation.\cite{square1} The valley index
 $\Xi$ denotes valley configuration of three-electron ground state in
 each sample. $S_{\rm tot}$ and ``Deg.'' represent the total spin and degeneracy
 factor of the three-electron ground state in each sample, respectively.
} 
\label{table1}
\end{table}

For a theoretical study, one can obtain the
total energy of multi-electron ground states and calculate the addition energy.
However, the well widths and effective diameters
required for quantitative calculation are unavailable in
Ref.\,\onlinecite{Borselli1}. Therefore, we first derive the well widths from
Eq.\,(\ref{eqA2}) with the reference value suggested by the experimental work
($d\sim 4$~nm).\cite{width} For Si3 and Si4, we use the valley splittings $\Delta
E^{\rm v}_{0}$ from the 
experiment.\cite{Borselli1} Since the valley splittings of Si1 and Si2 are
too small to be measured in experiment,\cite{Borselli1} we take them to be zero in our
calculation. Then, we calculate
the ground state energies of one-, two- and three-electron cases by treating $d_0$
as a parameter and determine its
value by the least square method of $\Delta \mu_1$ and $\Delta
\mu_2$.\cite{square1} In the previous section, we only introduce the frame of
the three-electron case. For the one-electron case, one needs to diagonalize
the single-electron Hamiltonian $H_e$ as given in Eq.\,(\ref{eq2}), while for
the two-electron case we follow the frame in Ref.\,\onlinecite{Lwang1}.

Our results of addition energies as well as the effective diameters of QDs are
listed in Table~\ref{table1}.
Good agreement with experimental data can be
observed for all devices with the largest relative error less than 3~\%, which
confirms the validity and accuracy of our model. In fact, one can also estimate the
effective diameter of the QDs solely from the fitting of $\Delta \mu_1$ with the
experimental value $\Delta \mu_1^\ast$. With the effective diameter obtained in
this way, we recalculate $\Delta \mu_2$ and find that the value
also agrees well with $\Delta \mu_2^\ast$ (within 4~\%). We should point
out that the Coulomb interaction here is very important in these quantum
systems. Without explicitly including the Coulomb interaction, the addition
energy is mainly from the orbital energy and the theoretical results become far
away from the experimental data. 

\begin{figure}
  \begin{center}
    \includegraphics[width=7cm]{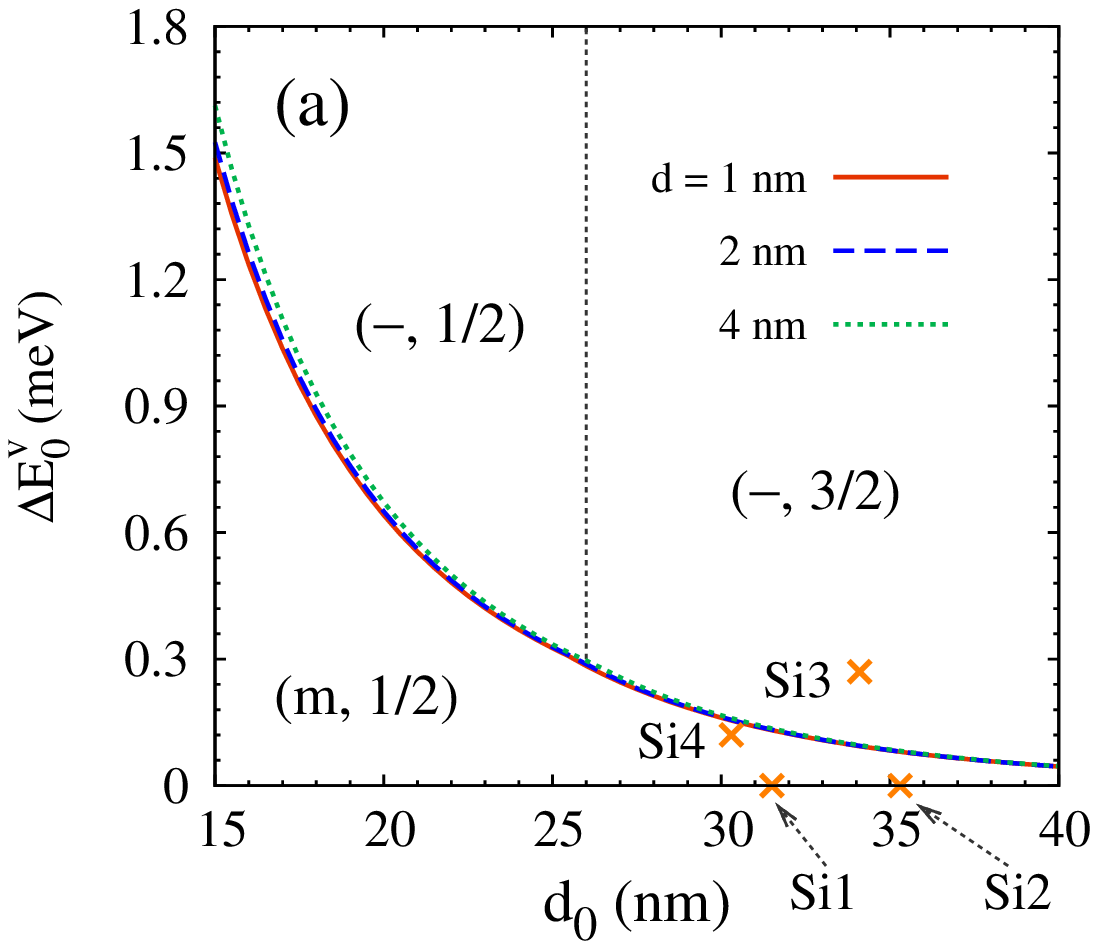}
    \includegraphics[width=7cm]{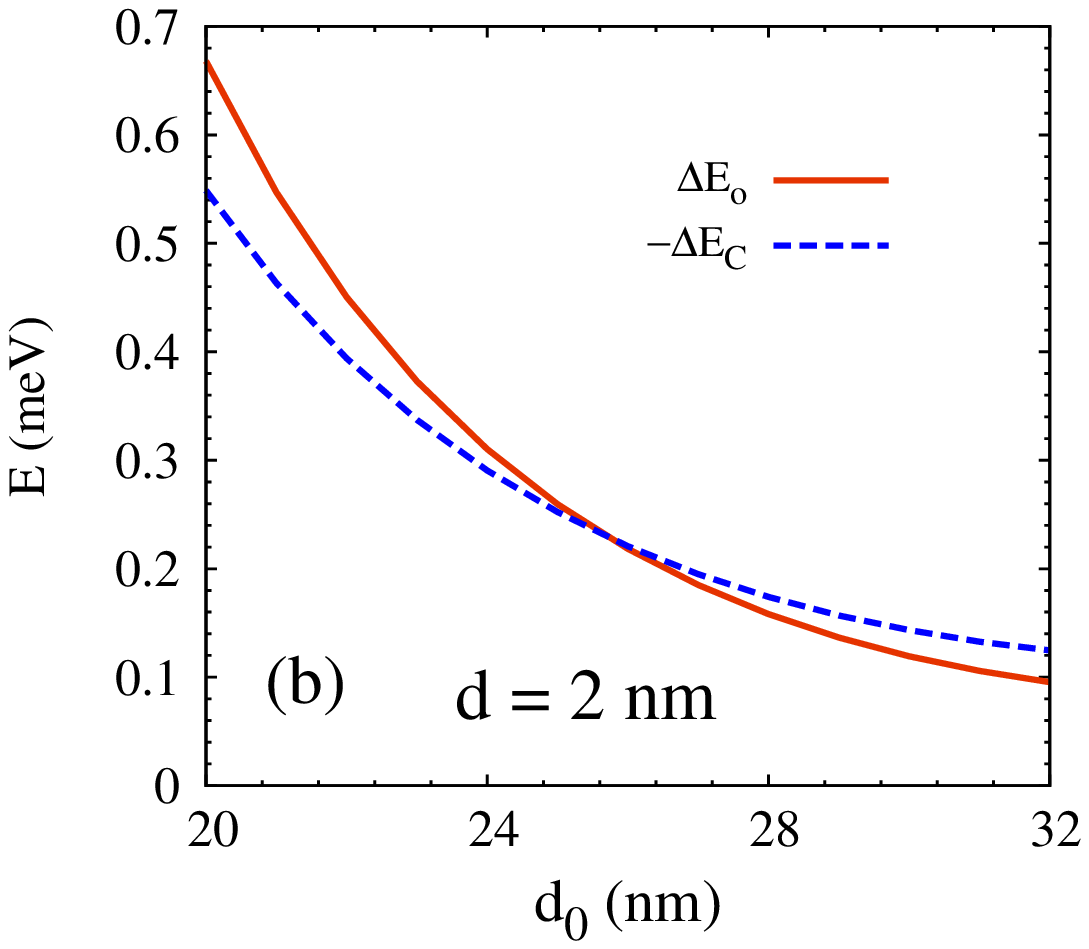}
  \end{center}
  \caption{(Color online) (a) Valley and spin configurations of the ground state
    in the absence of magnetic field. The curves represent the effective
    diameter dependence of the valley splitting
    energy for the emergence of the degeneracy between the lowest
    levels with ``$-$'' and ``m'' valley configurations.
    For each half-well width $d$,
    the ground state is of ``$-$'' (``m'') configuration above (below) the
    corresponding curve. The ``$-$''-configuration regime is separated into two parts
    by the vertical dotted line, according to the total spin $S_{\rm tot}$. The
    valley configuration and the
    total spin are labeled as ($\Xi$, $S_{\rm tot}$) in each regime.
    The crosses represent the parameters in Si1-Si4 in
    Ref.\,\onlinecite{Borselli1}. (b) The orbital 
    energy difference and the inverse number of Coulomb energy
    difference between the lowest quartet and doublet states with ``$-$''
    valley configuration {\em vs.}
    effective diameter for $d=2$~nm in the absence of the magnetic field.}
    \label{fig1}
\end{figure}
\subsubsection{Ground state configuration}
In Table~\ref{table1}, the valley configuration and total spin of the
three-electron ground states are also listed.
As shown in the table, the ground state in Si3 is mainly comprised
of the quartets with ``$-$'' valley configuration due to the large valley
splitting. This level is four-fold degenerate. However, the valley splitting in
Si4 is small and the ground state is two-fold degenerate doublet of
``m''-configuration. For Si1 and Si2, the states with ``$\widetilde{\rm m}$''
valley configuration are degenerate with those with
``m''-configuration. Therefore, the ground states are four-fold
degenerate. Since the states with different valley configurations are
decoupled as mentioned above,\cite{Dimi1,Lwang1} we explicitly investigate
the three-electron spectrum for each valley 
configuration individually in the following. The relative position between the spectra
  of different valley configurations is only determined by the
  exact value of the valley splitting in real system, which is also the
  criterion of the ground state configuration.

To elucidate the valley and spin configurations of the
ground state in the absence of the magnetic field, we draw a phase-diagram-like
picture, Fig.\,\ref{fig1}(a), by calculating the energy
difference $\Delta E_g^{\rm v}$ between the lowest level with pure (``$-$'') configuration
and that with mixed (``m'') configuration. From this figure, one finds three
regimes for each half-well width $d$, as labeled by
($\Xi$, $S_{\rm tot}$). It is seen that the ground state is of ``$-$'' (``m'') valley
configuration with large (small) valley splitting as expected. For the
``$-$''-configuration case, the ground
state can vary between quartet and doublet
as the dot size changes, with the transition occurring at around $d_0=26$~nm
(shown as the vertical dotted line). Since the energy of the lowest ``$+$'' (``$\widetilde
{\rm m}$'') configuration state is always higher than that of the lowest ``$-$'' (``m'')
configuration one due to a finite valley splitting, the ground state can not be of
this configuration. However, when $\Delta E^{\rm
  v}_0=0$, the lowest ``$\widetilde
{\rm m}$''-configuration level is degenerate with ``m''-configuration
one, therefore, can also be the ground state in that case.

In our calculation, we determine the borderline
between different valley configurations from the degenerate condition between them.
Specifically, we calculate the energy of the lowest ``$-$'' and ``m'' levels in
the absence of the valley splitting,
i.e., $E^{\rm p}_{\rm g}$ and $E^{\rm  m}_{\rm g}$ separately, 
and obtain $\Delta E^{\rm v}_{\rm g}=E^{\rm p}_{\rm 
  g}-E^{\rm m}_{\rm g}-\Delta E^{\rm v}_0$. The valley splitting on the
borderline is then given by $\Delta
\widetilde E^{\rm v}_0=E^{\rm p}_{\rm 
  g}-E^{\rm m}_{\rm g}$. Actually, the
real value of the valley splitting $\Delta E_0^{\rm v}$ for certain well width $d$ should be
determined by $\Delta_{0,0}^1$ in Appendix~\ref{appA}. If the coordinate
$(d_0,\Delta E_0^{\rm v})$ locates
above the curve of the corresponding well width, i.e., $\Delta E_0^{\rm v}>\Delta
\widetilde E^{\rm v}_0$ or $\Delta E^{\rm v}_{\rm g}<0$, the
ground state is of ``$-$'' valley configuration. Otherwise,
it should be of ``m''-configuration.

The origination of the variation of the ground state spin configuration
in the ``$-$''-configuration regime
is found to be the competition between the orbital and Coulomb
energies. The relative
orbital (Coulomb) energy between the lowest
quartet and doublet $\Delta E_{\rm o}$ ($\Delta E_C$)
is defined as the orbital (Coulomb) energy of the lowest
quartet subtracting that of lowest doublet. One calculates the orbital energy
from the expectation of $\sum_{i=1}^{3}H_e^{i}$ of the three-electron eigenstates, while the
Coulomb energy is similarly given by $H_C^{12}+H_C^{13}+H_C^{23}$. 
From the explicit expressions of the quartet and doublet given in
Appendix~\ref{appB}, one notices that all three electrons in
quartet must occupy different orbits while two electrons in
doublet can stay in the same one. Therefore, the lowest quartet
states have higher orbital energy than the lowest doublet states ($\Delta E_{\rm
  o}>0$). We finds that 
the Coulomb interaction of the lowest quartet states is smaller than that of the
lowest doublet states ($\Delta E_{C}<0$). In Fig.~\ref{fig1}(b), we plot
$\Delta E_{\rm o}$ and $-\Delta E_C$ as function of effective diameter, where an
intersecting between these two quantities can be seen. As a qualitative
understanding, $\Delta E_{\rm o}\propto 1/d^2_0$ can be characterized from
$\hbar\omega_0=\hbar^2\pi/(m_td^2_0)$, while $\Delta E_C\propto 1/d_0$ is estimated from
$e^2/(4\pi\epsilon_0\kappa d_0)$. Therefore, the orbital energy is more
sensitive to the diameter than the Coulomb interaction. For a small diameter QD,
the relative position between the lowest doublet and quartet is dominated by the
orbital energy and the ground state is doublet. As $d_0$ increases, $\Delta E_{\rm o}$
can become smaller than $-\Delta E_C$ with the
crossover at $d_0\sim 26$~nm in Fig.~\ref{fig1}(b), resulting in the
transition of the ground state from doublet to
quartet.

From Fig.~\ref{fig1}(a), one can directly read the ground state
configuration with the knowledge of the dot size and valley splitting, e.g.,
Si1-Si4 shown as crosses. 
Moreover, one notices that our phase-diagram-like picture is robust against
the well width, resulting from the strong confinement regime along $z$-direction
($d\ll d_0$).

\subsubsection{Three-electron spectrum with ``$-$'' valley configuration}
\begin{figure}
  \begin{center}
    \includegraphics[width=7.cm]{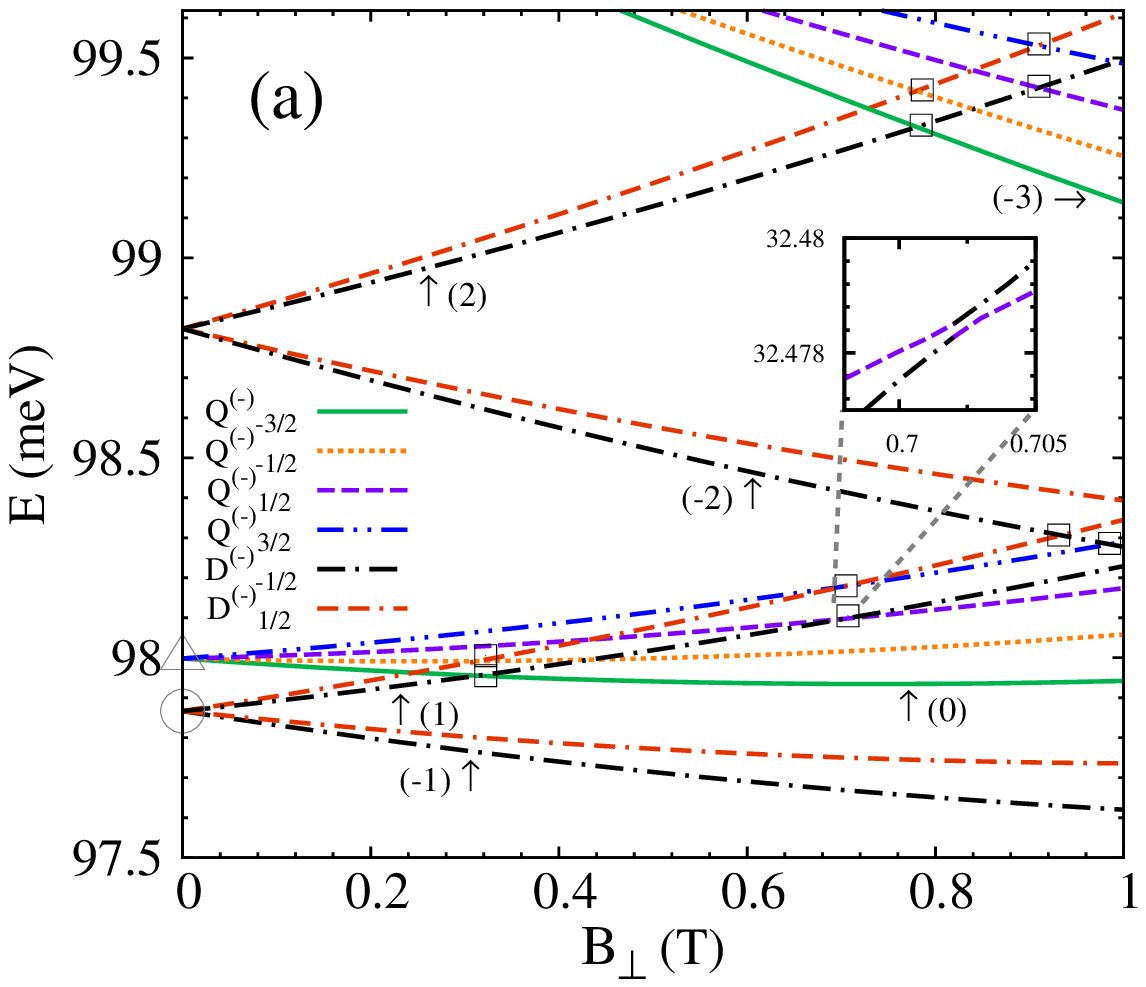}
    \includegraphics[width=7.cm]{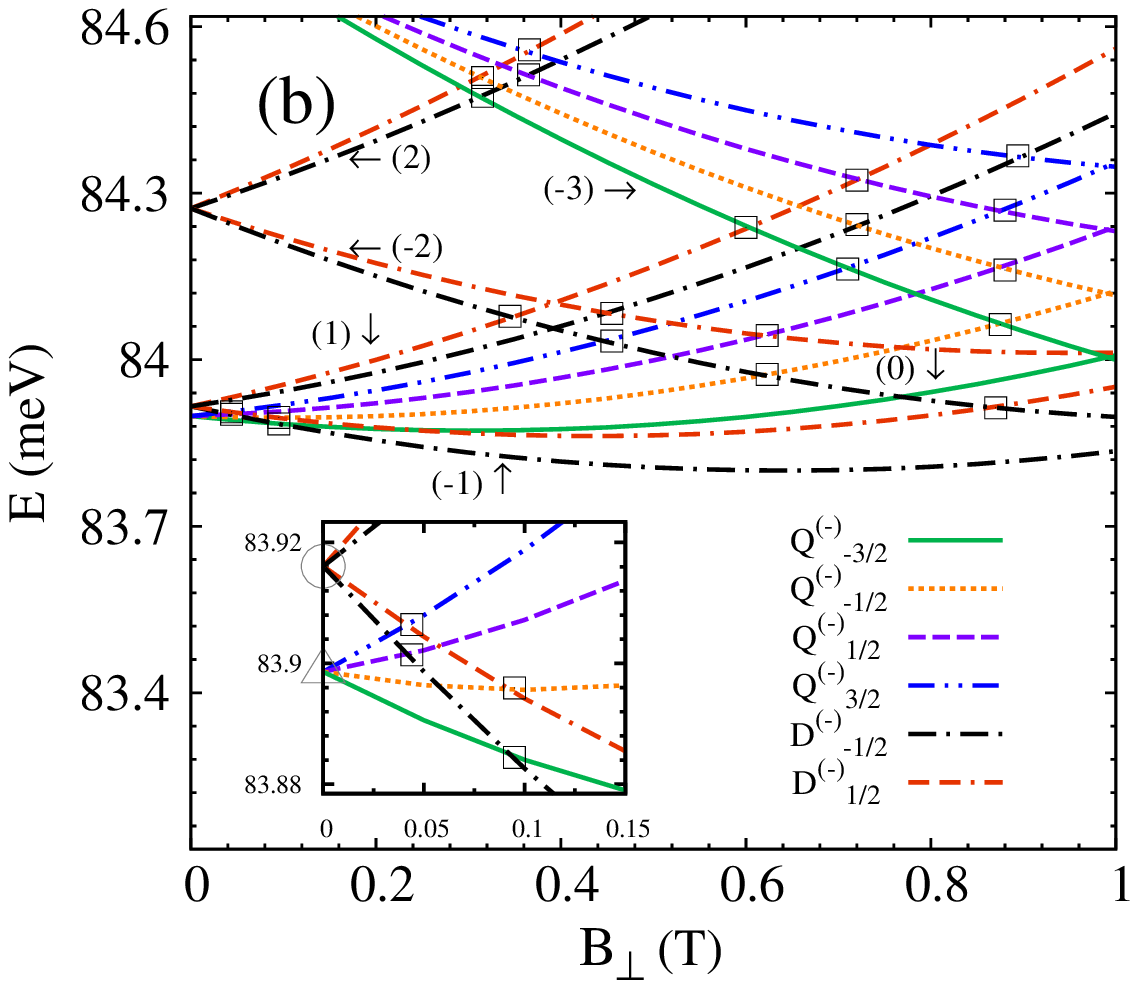}
  \end{center}
  \caption{(Color online) The lowest few energy levels {\em vs.} perpendicular
    magnetic field $B_{\perp}$ in single QDs with $d=2.12$~nm and the effective
    diameter $d_0=20$~nm (a) and 29~nm (b). The anticrossing points are labeled
    by open squares. The circle and open triangle at $B_\perp =0$ describe the lowest
    doublet and quartet states separately. 
    The total orbital angular momentum $L=l_1+l_2+l_3$ of the energy
    level are labeled correspondingly.\cite{orbit} The
    inset in (a) zooms the vicinity of the
    anticrossing point at $B_{\perp}\sim 0.72$~T. The inset in (b) enlarges the
    spectrum in the vicinity of $B\sim 0$~T. }
    \label{fig2}
\end{figure}

In this part, we take advantage of
our model to calculate the three-electron energy spectrum with ``$-$''-valley
configuration. We study the perpendicular magnetic field dependence at
$d=2.12$~nm (corresponding to $\Delta E_0^{\rm v}\approx 1.9$~meV) with
$d_0=20$~nm and $29$~nm, where the ground states lie in the
$(-,1/2)$ and $(-,3/2)$ configuration regimes, respectively. The lowest few 
energy levels are plotted as function of
$B_{\perp}$ in Fig.~\ref{fig2}. These eigenstates are
denoted as $Q^{(-)}_{-3/2}$, $Q^{(-)}_{-1/2}$, $Q^{(-)}_{1/2}$,
$Q^{(-)}_{3/2}$, $D^{(-)}_{-1/2}$ and $D^{(-)}_{1/2}$, according to the total
spin $S_{\rm tot}$ and $S_z$ of the major components in these
states. The lowest doublet (quartet) states at $B_{\perp}=0$~T are labeled as
circle (open triangle).
It is seen that the ground state for $d_0=20$~nm is four-fold degenerate
doublet, while that for $d_0=29$~nm is quartet, consistent with
  Fig.\,\ref{fig1}(a). As the magnetic field 
increases, the degeneracy is lifted due to the effect of the magnetic
field on Landau level and Zeeman splitting, resulting in many intersecting
points as shown in Fig.~\ref{fig2}(a) and (b).
We find that some intersecting points show anticrossing behavior
due to the SOCs.\cite{Lwang1} In this work, all the anticrossing points are
labeled by open squares. In the inset, we enlarge the spectrum in the vicinity
of one anticrossing point at $B_{\perp}\sim 0.72$~T, where an energy gap
$\sim 0.24$~$\mu$eV is present.\cite{Rashba1,Vervoort1,Vervoort2} 
As reported, the anticrossing points are important for spin
manipulation, due to the strong spin mixed at these
points.\cite{Petta1,Nowack1,Sarkka1,Laird1,Petta2,Lwang1,Lwang2}

Since the eigenstates are strongly mixed, e.g., the largest
component in the ground states is usually in the order of 20\%, one can not
naively describe an eigenstate by a single basis function. However, away from
the anticrossing points, the SOCs are weak and the
total orbital angular momentum $L=l_1+l_2+l_3$ is still a good quantum
number. In Fig.~\ref{fig2}(a), we labeled this quantum number for each state.
With this quantum number, one can understand why an intersecting point is
anticrossing or not in the presence of SOCs. By rewriting the SOCs in the form of ladder
operators, one obtains
$H_{\rm so} = \tfrac{2ia_0}{\hbar}(P^+S^--P^-S^+)-\tfrac{2b_0}{\hbar}(P^+S^++P^-S^-)$
with $P^{\pm}=(P_x\pm iP_y)/2$ and $S^{\pm}=S_x\pm iS_y$. Since $P^{\pm}$
($S^{\pm}$) changes $L$ ($S_z$) by one unit, the states with
($L$, $S_z$) can only couple with the states with ($L\mp 1$, $S_z\pm 1$) via
Rashba term and those with ($L\pm 1$, $S_z\pm 1$) via IIA
term.\cite{Lwang1} For example, in the inset of Fig.~\ref{fig2}(a), the
spin-down doublet with ($1$, $-\tfrac{1}{2}$) (black chain curve) couples with
the quartet with ($0$, $\tfrac{1}{2}$) (purple dashed curve)
via IIA term, resulting in the anticrossing behavior.

\begin{figure}
  \begin{center}
    \includegraphics[width=6.5cm]{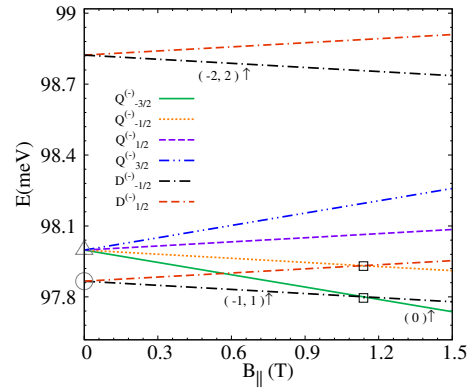}
  \end{center}
  \caption{(Color online) The lowest few energy
    levels {\em vs.} parallel magnetic field $B_{\parallel}$ in single QDs.
    Open squares indicate the anticrossing points. The 
    circle and open triangle at $B_\perp=0$ describe the lowest doublet and quartet in the
    absence of the magnetic field. The total orbital angular momentum
    $L=l_1+l_2+l_3$ are labeled. The half-well width
    $d=2.12$~nm and the effective diameter $d_0=20$~nm.}
  \label{fig3}
\end{figure}

To exclude the orbital effect, we apply a parallel magnetic field instead.
The results with $d=2.12$~nm and $d_0=20$~nm are shown
in Fig.~\ref{fig3}, where
the eigenstates are denoted as $Q^{(-)}_{-3/2}$, $Q^{(-)}_{-1/2}$, $Q^{(-)}_{1/2}$,
$Q^{(-)}_{3/2}$, $D^{(-)}_{-1/2}$ and $D^{(-)}_{1/2}$  according to
$S_{\rm tot}$ and the $x$-direction component $S_x$ of
their major components. In this case, the energy levels are linearly dependent on
the magnetic field due to the Zeeman splitting.
The degeneracy from the orbital degree of freedom survives. For example,
the ground state for $B_\|<1$~T is two-fold degenerate with $L=\pm 1$ as
shown in Fig.\,\ref{fig3}. Similar to the situation with the perpendicular
magnetic field, we also observe anticrossing points, e.g., the ones marked as
open squares at $B_{\parallel}\sim 1.1$~T. However, we should point out that
the criterion of the anticrossing points here are different from that with
perpendicular magnetic field.\cite{Lwang1} In this case, the SOCs can be written as
$H_{\rm so}=\left[{a_0}(P^++P^-)-ib_0(P^+-P^-)\right](\widetilde S^++\widetilde
S^-)/\hbar$ with $\widetilde S^\pm =S_y\pm iS_z$.\cite{Lwang1}

\subsubsection{Three-electron spectrum with ``m'' valley  configuration}
\begin{figure}
  \begin{center}
    \includegraphics[width=7.cm]{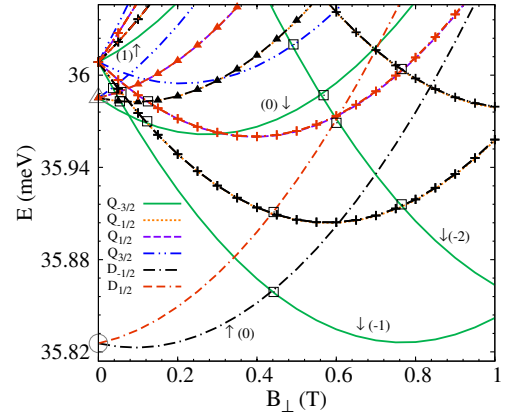}
  \end{center}
  \caption{(Color online) The lowest few energy
    levels with the ``m'' valley configuration {\em vs.} perpendicular
      magnetic field $B_{\perp}$ in 
      single QDs. Open squares indicate the anticrossing points. The
    curves with crosslets denote the three-fold degenerate levels (two
    doublet states and one quartet state) and the ones with solid triangles
    represent the two-fold degenerate levels (one doublet state and one
    quartet state). The circle and open triangle
    at $B_\perp=0$ describe the lowest doublet and quartet and doublet. The
    total orbital angular momentums
    $L=l_1+l_2+l_3$ are labeled. The half-well width
    $d=3.926$~nm and the effective diameter $d_0=30.3$~nm.}
  \label{fig4}
\end{figure}

As shown in Table~\ref{table1} and Fig.\,\ref{fig1}, the valley configuration of
the ground state can be ``m'' for small valley splitting case, e.g., in
Si4 with $\sim0.12$ $\mu$eV. In this part, we study the
energy spectrum with the ``m'' valley configuration to illustrate the role of
the valley degree of freedom in the three-electron spectrum. In Fig.~\ref{fig4}, we plot the
perpendicular magnetic field dependence of the energy spectrum with
the ``m'' valley configuration
in Si4 ($d=3.926$~nm and $d_0=30.3$~nm). Similarly, the lowest
several eigenstates here are
denoted as $Q^{(-)}_{-3/2}$, $Q^{(-)}_{-1/2}$, $Q^{(-)}_{1/2}$,
$Q^{(-)}_{3/2}$, $D^{(-)}_{-1/2}$ and $D^{(-)}_{1/2}$, according to the total
spin $S_{\rm tot}$ and $S_z$ of the major components.
The lowest doublet and quartet states at $B_{\perp}=0$~T are labeled as
red circle and open triangle, respectively. In the ``m'' valley configuration case, each
single-electron orbit (distincted by the quantum numbers $n$, $l$ and $n_z$) is
four-fold degenerate due to the spin and valley degrees of freedom. Therefore,
at most two
electrons in quartet can stay in the same single-electron orbit, while
all the three electrons in doublets can stay in the same
one. This gives rise to the
difference between energy spectrum of the ``m'' and ``$-$'' valley
configurations.

Interestingly, we find that the magnetic-field independent degeneracy
between quartet and 
doublet can exists in the ``m'' valley configuration. We notice that the
degeneracy due to the valley degree of freedom has been discussed by Wang 
{\em et al.} in two-electron case in Ref.\,\onlinecite{Lwang1}. Here, we
find that the present degeneracy can be either two-fold or three-fold.
In Fig.~\ref{fig4}, the three-fold degenerate levels (one quartet state
and two doublet states) are labeled by crosslets, while the
two-fold degenerate ones (one quartet state and one
doublet state) are marked by solid triangles. Actually, this energy-level
degeneracy mainly results from the negligibly small
intervalley Coulomb interaction and SOCs.\cite{Dimi1,Lwang1}
One finds that the doublet states in the two-fold degenerate case are mainly
comprised of $|D^{(m)(2)}_{\pm1/2}\rangle$ elements.
This is associated with the situation that 
$|D^{(m)(2)}_{\pm1/2}\rangle$ basis functions of the ``m'' configuration are
decoupled with $|D^{(m)(1)}_{\pm1/2}\rangle$ and $|D^{(m)(3)}_{\pm1/2}\rangle$
in the absence of the intervalley Coulomb interaction and SOCs. 
Moreover, one finds that there is one $|Q^{(m)}_{\pm1/2}\rangle$ basis
function with the same orbital construction as each
$|D^{(m)(2)}_{\pm1/2}\rangle$ and vice verse, according to Appendix~\ref{appB}.\cite{nv}
The corresponding Hamiltonian elements for these two sets are equal, i.e.,
\begin{eqnarray}
  \nonumber &&  \langle D^{(m)(2)}_{\pm1/2} |H_{\rm
    tot}|D^{(m)(2)'}_{\pm1/2}\rangle=   \langle Q^{(m)}_{\pm1/2} |H_{\rm
    tot}|Q^{(m)'}_{\pm1/2}\rangle\\
  \nonumber
  &=&\langle N_1N_2N_3 |H_{\rm
    tot}|N'_1N'_2N'_3\rangle \pm \tfrac{1}{2}g\mu_B B_z\\
  &&  -\langle
  N_1N_2N_3 |H_{\rm tot}|N'_2N'_1N'_3\rangle 
  \label{eq10}
\end{eqnarray}
Thus it's clear that each quartet energy level with $S_z=\pm\frac{1}{2}$
degenerates with one doublet
state (constructed by $|D^{(m)(2)}_{\pm1/2}\rangle$).

For the three-fold degenerate levels shown as the
curves with crosslets in Fig.~\ref{fig4}, we find the additional doublet state is
combined by $|D^{(m)(1)}_{\pm1/2}\rangle$ and
$|D^{(m)(3)}_{\pm1/2}\rangle$ basis functions. However, the combination of these
basis functions is very complex, therefore, we skip the discussion in this work.

\subsection{Vertically coupled double quantum dots}

\begin{figure}[htb]
  \begin{minipage}[]{10cm}
    \hspace{-1.5 cm}\parbox[t]{5cm}{
      \includegraphics[height=4.5 cm]{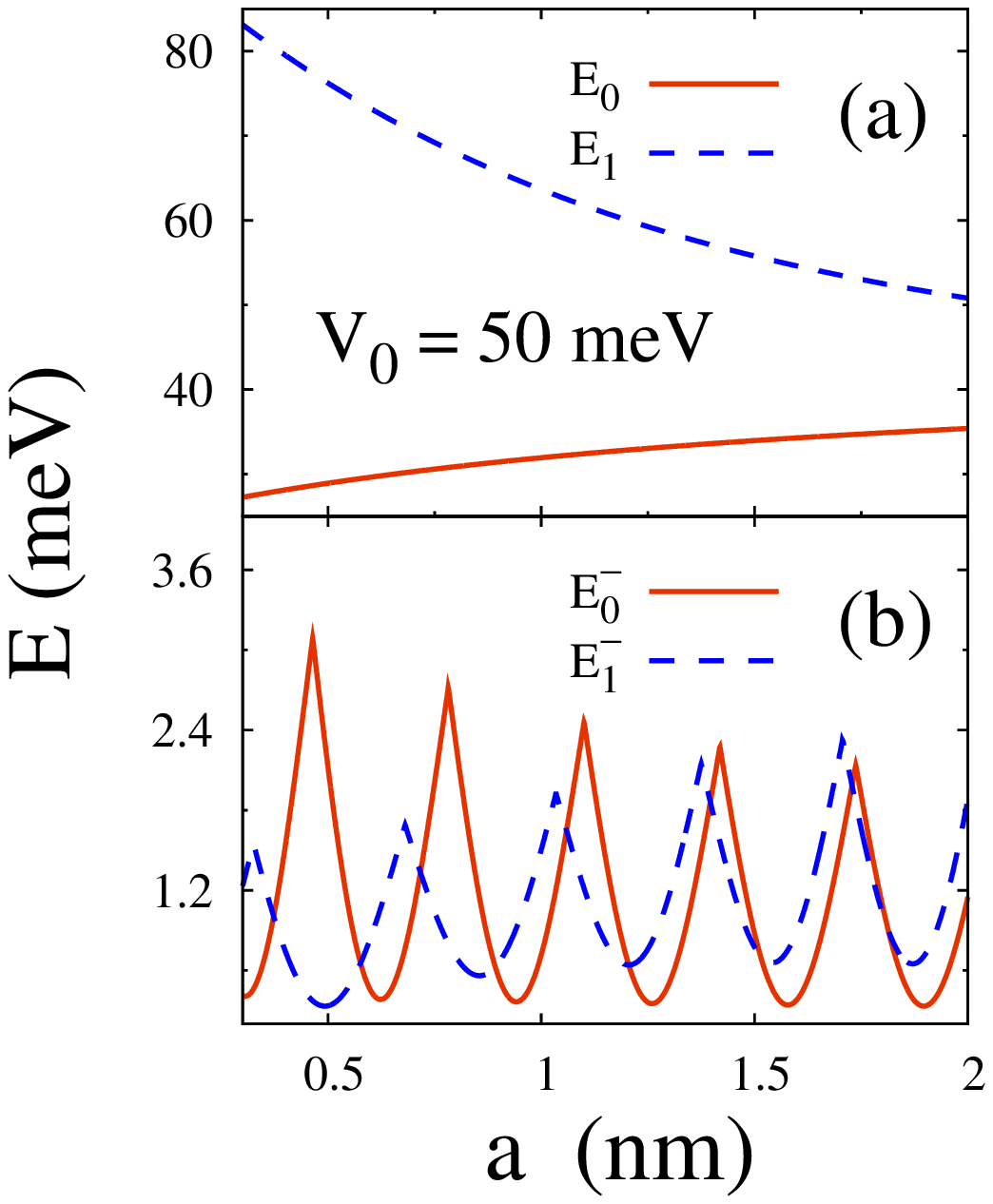}}
    \hspace{-0.7 cm}\parbox[t]{5cm}{
      \includegraphics[width=4.5cm]{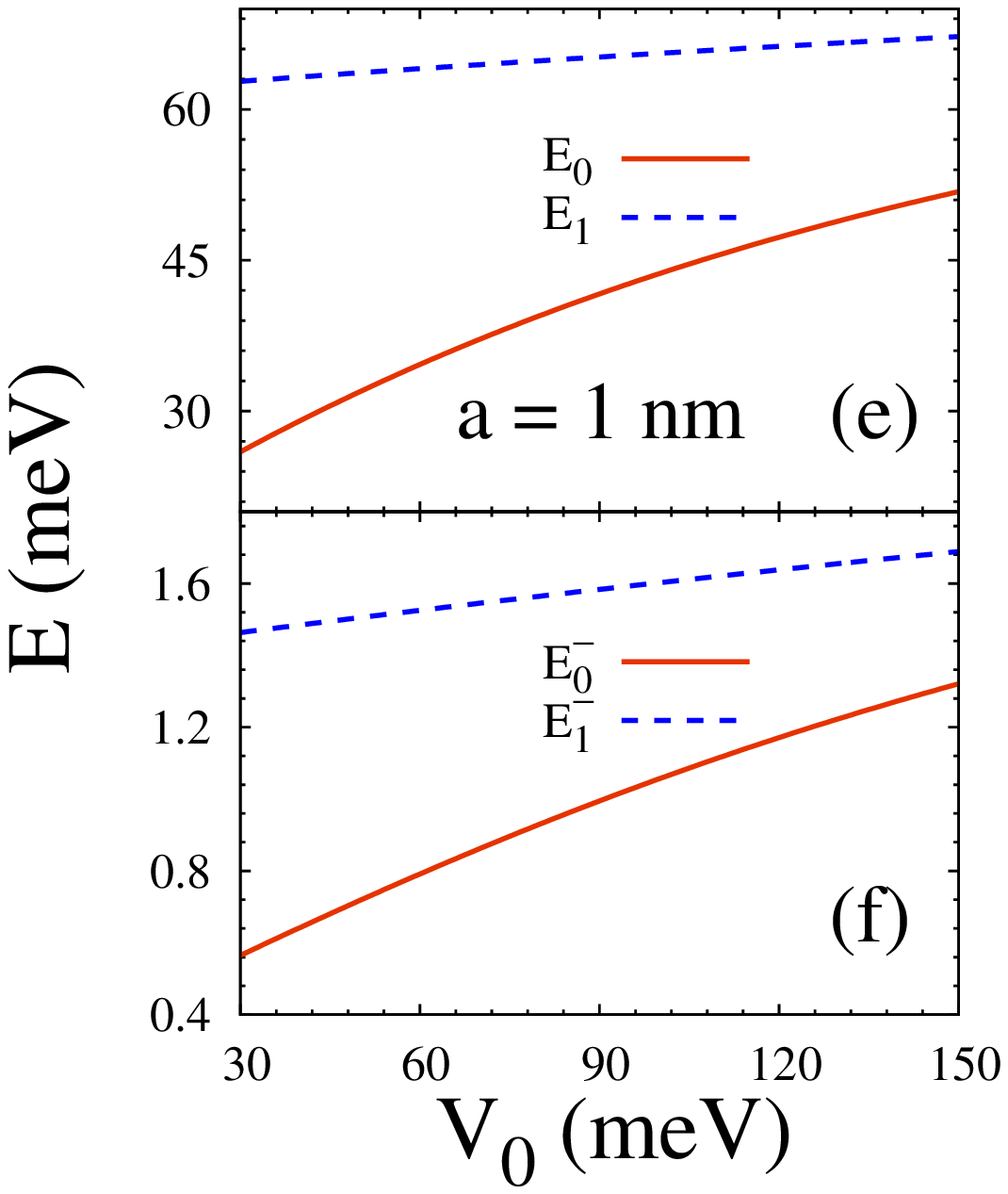}}
  \end{minipage}
  \begin{minipage}[]{10cm}
    \hspace{-1.5 cm}\parbox[t]{5cm}{
      \includegraphics[width=4.5cm,height=4.5 cm]{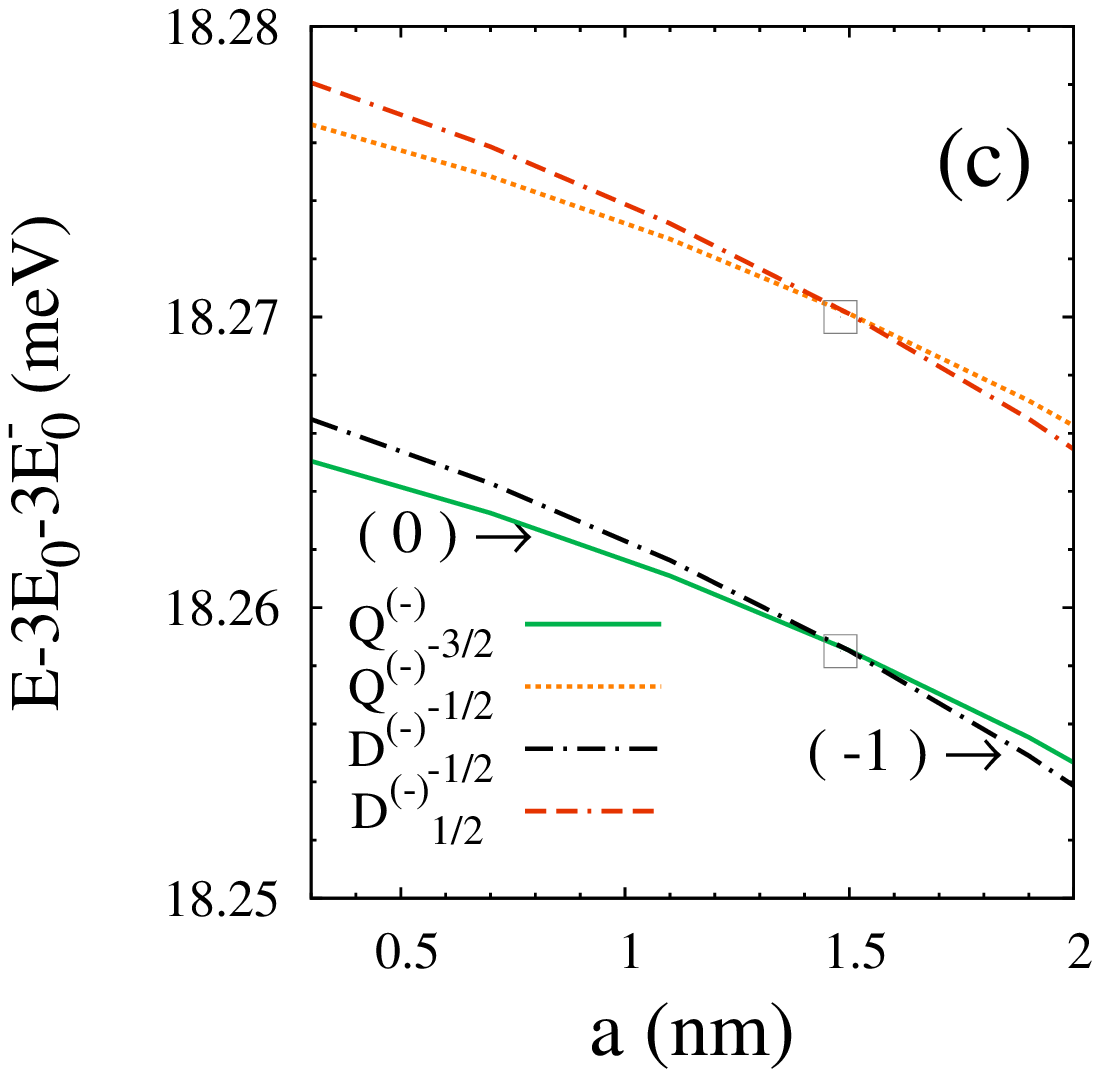}}
    \hspace{-0.7 cm}\parbox[t]{5cm}{
      \includegraphics[width=4.5cm,height=4.5 cm]{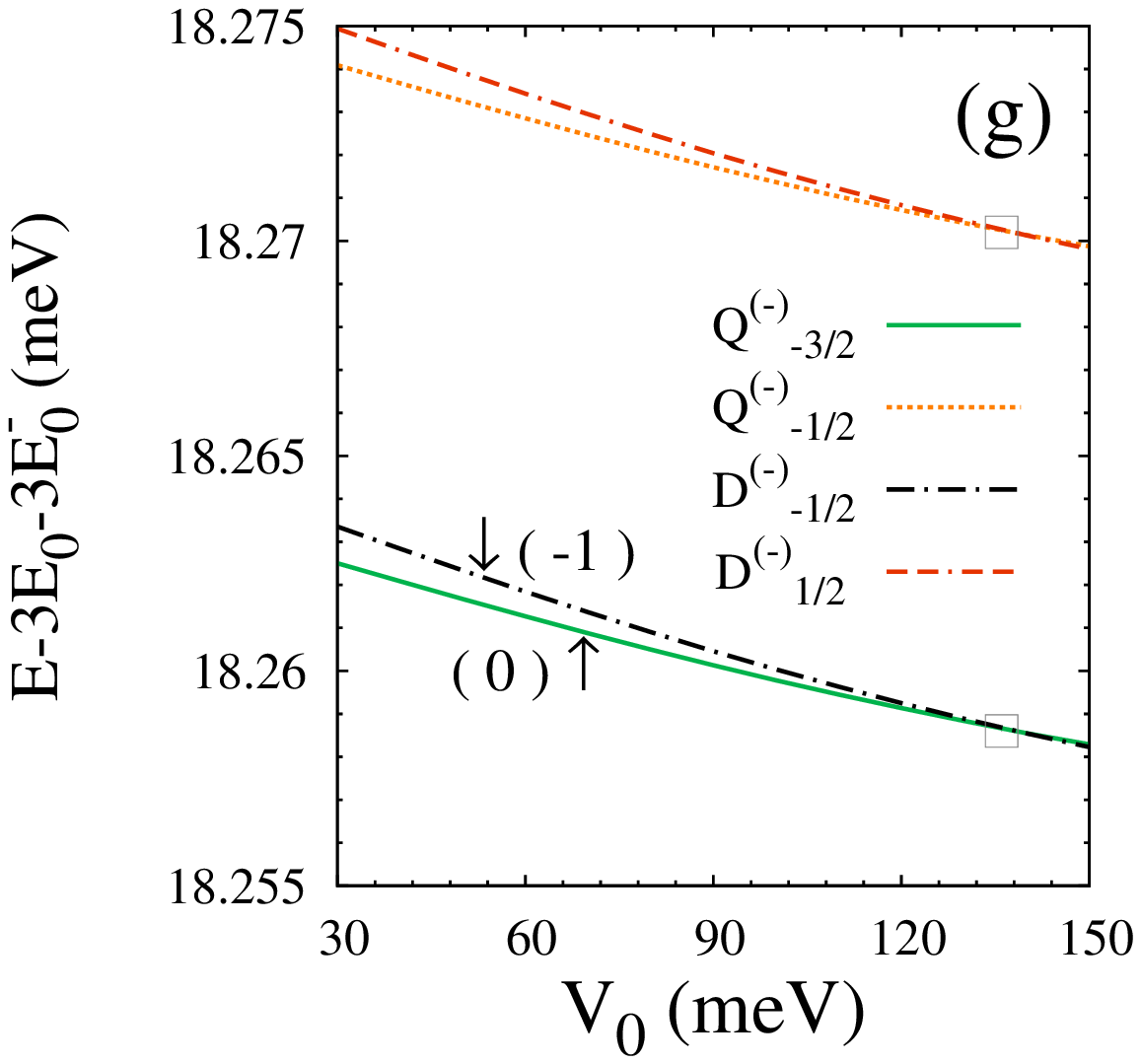}}
  \end{minipage}
    \begin{minipage}[]{10cm}
      \hspace{-1.5 cm}\parbox[t]{5cm}{
        \includegraphics[width=4.5cm,height=4.5 cm]{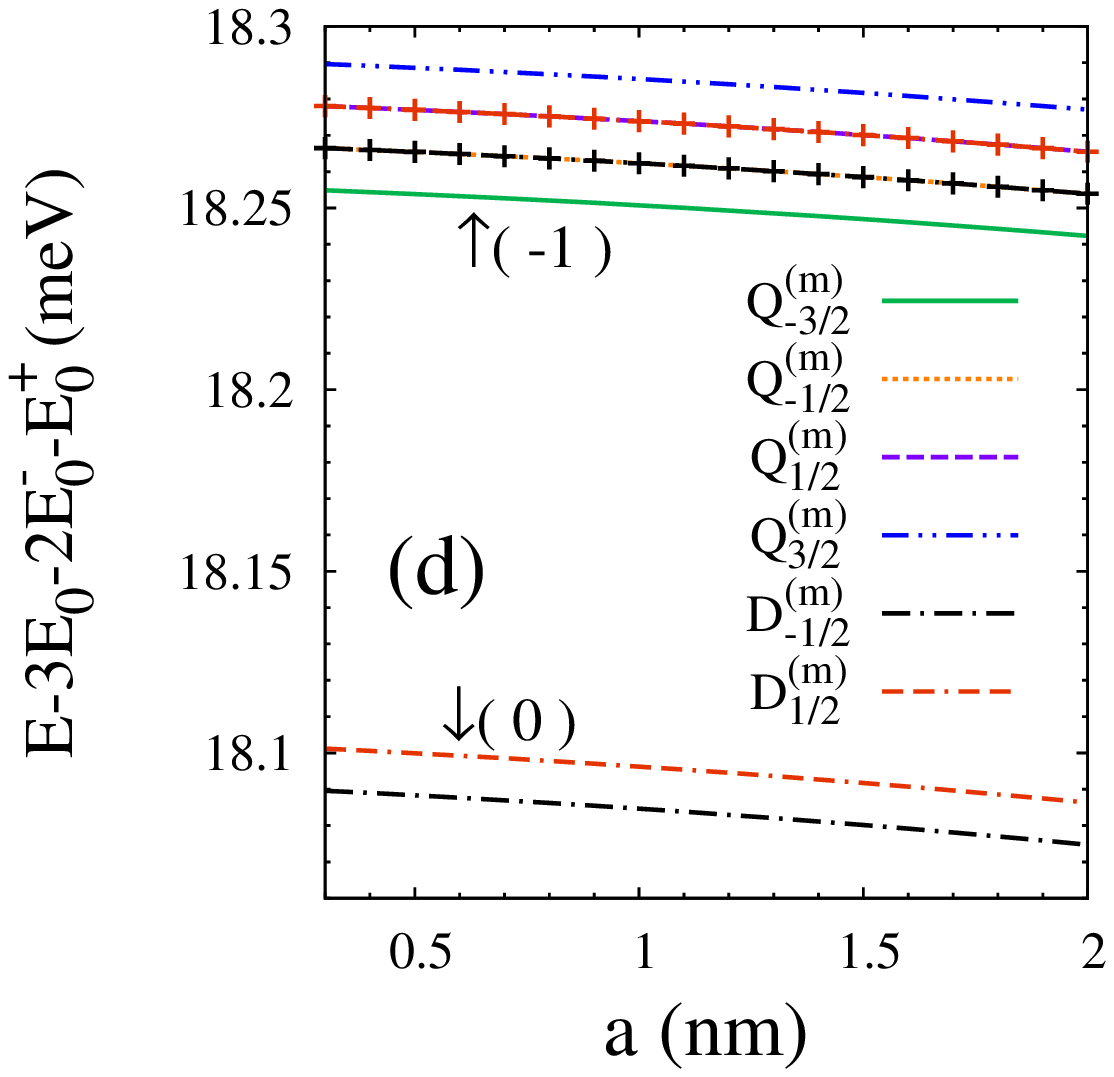}}
      \hspace{-0.7 cm}\parbox[t]{5cm}{
        \includegraphics[width=4.5cm,height=4.5 cm]{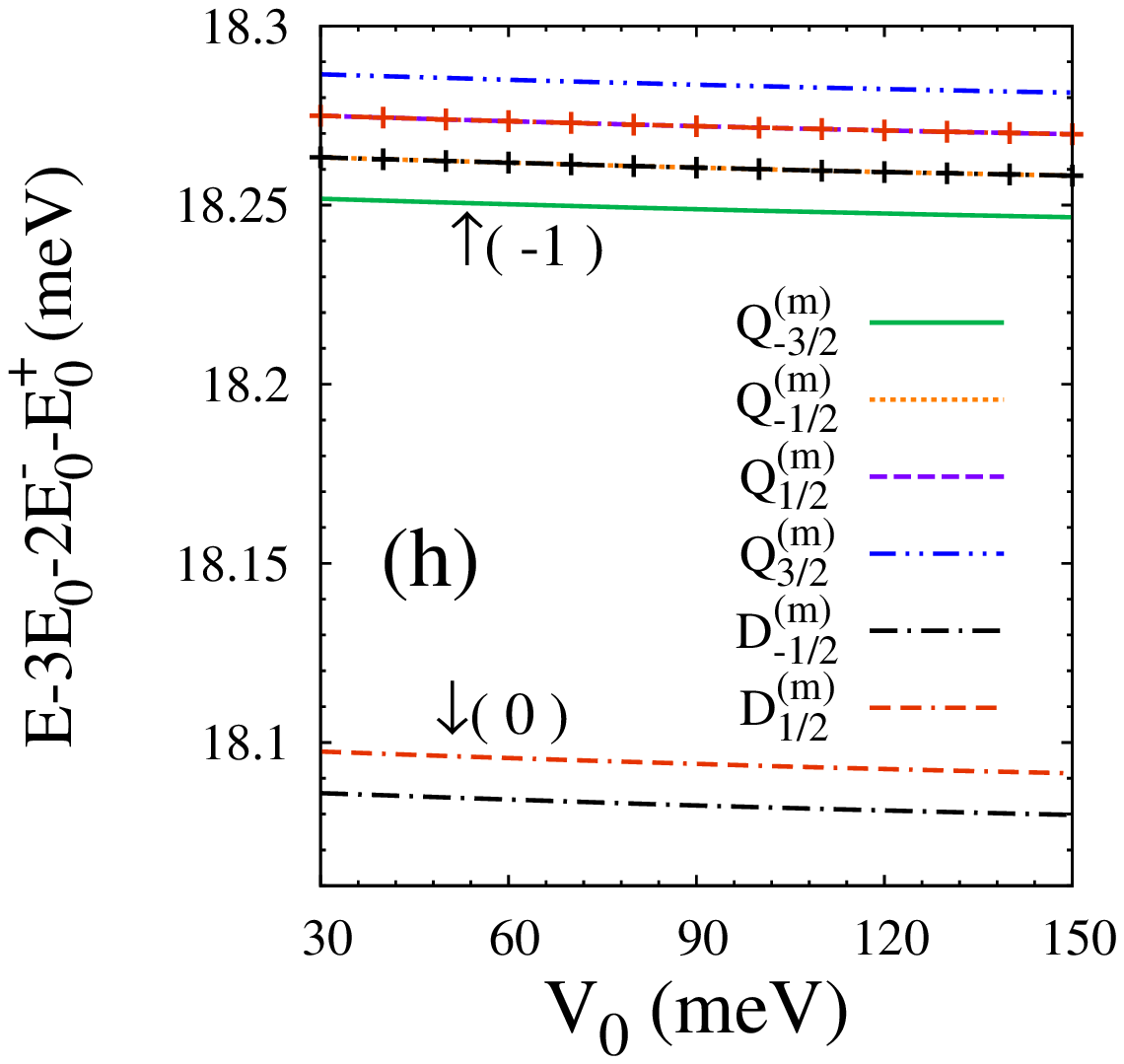}}
    \end{minipage}
   \caption{(Color online) (a) and (e) represent the interdot barrier-width and
     barrier-height dependences of the lowest two subband energies along $z$-direction.
     (b) and (f) are the valley energies of the lower valley
    eigenvalue of the lowest two subbands as function of the interdot barrier
    width and barrier height. (c) and (g) show the three-electron energy
    spectrum with ``$-$'' valley configuration, while (d) and (h) give
    the corresponding spectrum with ``m'' valley configuration.
    Open squares in (c) and (g) indicate the anticrossing points. The
    curves with crosslets denote the three-fold degeneracy of the levels (two
    doublet states and one quartet state). The total orbital angular momentum
    $L=l_1+l_2+l_3$ of each states are labeled. Here, the effective diameter
    $d_0=29$~nm, the well width $d=2$~nm and $B_{\perp}=0.1$~T. In (a)-(d),
    the barrier height $V_0=50$~meV. In (e)-(h), the barrier width $a=1$~nm.} 
  \label{fig5}
\end{figure}

In this part, we investigate the vertically coupled DQD case with a
perpendicular magnetic field $B_{\perp}=0.1$~T with both the ``$-$'' and ``m'' valley
configurations. We first study the
interdot barrier-width dependence of the ``$-$'' valley configuration.
In Fig.~\ref{fig5}(a), the orbital energies of the lowest two subbands along the
$z$-direction are plotted as function of interdot barrier width. One can see that
the orbital energy of the first subband increases while that of the second subband
decreases as the barrier width increases, resulting the decrease of the energy
difference $E_1-E_0$ between them. This corresponds the decrease of
the interdot coupling. Differently, the valley energies of the lowest two subbands
present oscillation against the barrier width, which can be expected from the
formula given in Appendix~\ref{appA}. 
Since the intersubband coupling due to
the valley-orbit coupling is much weaker than
the intersubband splitting from the orbital degree of freedom,
we neglect this term in our calculation. Actually, the first subband dominates
the lowest few states because of the large intersubband
splitting ($>10$~meV) in the strong interdot coupling regime studied here.
The lowest four levels are plotted in Fig.\,\ref{fig5}(c), where the energies from
the subband and valley are subtracted for improving the solution of levels. Similar to the
single QD case, the quartet-doublet transition of the ground state can occur
and the anticrossing behavior (illustrated as open squares at $a\sim 1.5$~nm) can
be observed. We also study the barrier-height dependence with ``$-$''
  valley configuration and plot the results as
Fig.~\ref{fig5}(e)-(g). In Fig.~\ref{fig5}(g), the quartet-doublet
transition of the ground state and the anticrossing properties are also present.

As a comparison, we also show the energy spectrum of the ``m'' valley
configuration case as Fig.~\ref{fig5}(d) and (h), where the ground state is
doublet. In these figures, the curves with crosslets also represent the three-fold
degeneracy (two doublet state and one 
quartet state).

\section{SUMMARY}
In summary, we have studied three-electron energy spectra in Si/SiGe
single QDs and vertically coupled DQDs by using
the exact-diagonalization method. In our calculation, the Zeeman splitting, SOC,
valley coupling and electron-electron
Coulomb interaction are 
explicitly included. Due to the strong Coulomb interaction, a large
number of basis functions are employed to converge the energy
spectrum. The ground state energies in single QDs show good agreement
with the experiment. As a supplement of the experimental data, we identify the
valley configuration, spin configuration and the degeneracy factor of the ground state
from our calculation. We then systematically study the ground state
configuration in the absence of magnetic field, and find that the ground state
is of pure and mixed valley configurations with large and small valley
splittings, respectively. We show that the ground state with mixed valley
configuration is always doublet. In contrast, the ground state with pure valley
configuration is doublet in small dots, while it can be quartet in large dots.
Then, we explicitly study the energy spectra of the
three-electron states with two typical valley configurations, i.e., pure and
mixed valley states, in the
presence of perpendicular and parallel magnetic fields. In
the pure valley configuration case, we show that the doublet-quartet
transition of the ground state can be realized by tuning the magnetic field and/or
dot size. For the mixed valley configuration, the doublet-quartet transition of
the ground state can also be realized by magnetic field and
the three-electron energy spectrum
presents interesting quartet-doublet degeneracy,
which is connected with the negligible intervalley
coupling. Due to the spin-orbit coupling, the intersecting points between
the energy levels with identical valley configuration but different spin quantum
numbers can present anticrossing
behavior, which is expected to benefit
the manipulation of spin. We point out and analyze all these anticrossing points
in each case. Furthermore, we study the barrier-width and barrier-height
dependences of the three-electron energy spectrum in vertically strong-coupled DQDs.
We also find anticrossing points in the pure valley configuration case and
quartet-doublet degeneracy in the mixed configuration case.

\begin{acknowledgments}
We would like to thank M. W. Wu for proposing the topic as well
as directions during the investigation. One of the authors (Z.L.)
would also like to thank Y. Yin for valuable discussions. This work
was supported by the National Natural Science
Foundation of China under Grant No.\,10725417 and
the National Basic Research Program of China under
Grant No.\,2012CB922002.
\end{acknowledgments}

\begin{appendix}
  \section{VALLEY COUPLING ELEMENTS}\label{appA}
  \begin{eqnarray} 
    \Delta^0_{0,0}&=&\frac{V_{\rm v}\hbar^2{k_0}^2{C_0}^2}{m_z}+2V_{\rm v}V_0{C_0}^2\sin^2(k_0d),\\
    \nonumber  \Delta^1_{0,0}&=&\frac{V_{\rm
        v}\hbar^2{k_0}^2{C_0}^2\cos[2k_{\rm Si}(\frac{a}{2}+d)]}{m_z}
    \label{eqA2}
    \\
    &&\hspace{-0.3cm}{}+2V_{\rm v}V_0{C_0}^2\sin^2(k_0d)\cos(k_{\rm Si}a),\\
    \Delta^0_{1,1}&=&\frac{V_{\rm v}\hbar^2{k_1}^2{C_1}^2}{m_z}+2V_{\rm
      v}V_0{C_1}^2\sin^2(k_1d),\\
 \nonumber  \Delta^1_{1,1}&=&\frac{V_{\rm v}\hbar^2{k_1}^2{C_1}^2\cos(2k_{\rm Si}(\frac{a}{2}+d))}{m_z}\\
 &&\hspace{-0.3cm}{}+2V_{\rm v}V_0{C_1}^2\sin^2(k_1d)\cos(k_{\rm Si}a),\\
 \nonumber \Delta^1_{0,1}&=&-\Delta^1_{1,0}=i\{\frac{V_{\rm
     v}\hbar^2k_0k_1C_0C_1}{m_z}\sin[k_{\rm Si}(a+2d)]\\
 &&\hspace{-0.3cm}{}+ 2V_{\rm v}V_0C_1C_0\sin(k_0d)\sin(k_1d)\sin(k_{\rm Si}a)\},
\end{eqnarray}
where $V_{\rm v}=7.2\times10^{-11}$~m represents the ratio of the valley coupling strength to
the depth of the quantum well potential.\cite{Friesen1} 

\section{THREE-ELECTRON  BASIS FUNCTIONS}\label{appB}
The three-electron basis functions can be constructed in form of either doublet
($S_{\rm tot}=1/2$) or quartet ($S_{\rm tot}=3/2$). We first
combine the spin wavefunctions via Clebsch-Gordan coefficients.
For the doublet states, the spin functions can be expressed by 
\begin{eqnarray}
  \nonumber      \chi^\lambda_{1/2}&
  =&[(\beta_1\alpha_2+\alpha_1\beta_2)\alpha_3-2\alpha_1\alpha_2\beta_3]/{\sqrt{6}}\\[-2pt] 
  \nonumber      \chi^\lambda_{-1/2}&=&[2\beta_1\beta_2\alpha_3
  -(\beta_1\alpha_2+\alpha_1\beta_2)\beta_3]/{\sqrt{6}}\\[-2pt] 
  \nonumber      \chi^\rho_{1/2}& =&(\beta_1\alpha_2-\alpha_1\beta_2)\alpha_3/{\sqrt{2}}\\[-2pt]
  \chi^\rho_{-1/2}&=&(\beta_1\alpha_2-\alpha_1\beta_2)\beta_3/{\sqrt{2}},
  \label{B1}
\end{eqnarray}
and for the quartet states, one has
\begin{eqnarray}
  \nonumber     \chi^S_{3/2} &=&\alpha_1\alpha_2\alpha_3 \\[-2pt]
  \nonumber     \chi^S_{1/2}&=&(\alpha_1\alpha_2\beta_3+\alpha_1\beta_2\alpha_3
  +\beta_1\alpha_2\alpha_3)/{\sqrt{3}}\\[-2pt]
  \nonumber     \chi^S_{-1/2}& =&(\alpha_1\beta_2\beta_3+\beta_1\beta_2\alpha_3
  +\beta_1\alpha_2\beta_3)/{\sqrt{3}}\\[-2pt]
  \chi^S_{-3/2}& =&\beta_1\beta_2\beta_3.
  \label{B2}
\end{eqnarray}
Here, $\alpha_i=|\uparrow\rangle_i$ and
$\beta_i=|\downarrow\rangle_i$ with the subscript denoting the three electrons.
The superscript ``S'' labels the
symmetric configuration for the permutation of any two electrons, while
``$\lambda$'' and ``$\rho$'' represent the symmetric and antisymmetric
configurations for the permutation only for the electrons ``1'' and ``2'',
respectively. With these spin wavefunctions, we then add the corresponding orbital
parts to construct the total basis functions.
When the three electrons occupy three different orbits, i.e., $N_i\neq N_j$
with $i,j\in \{1,2,3\} $, one has the total wavefunctions of the doublet states
\begin{eqnarray}
  \nonumber |D^{(\Xi)(1)}_{1/2}\rangle&=&\phi^{\rho(1)}\chi^\lambda_{1/2}
  -\phi^{\lambda(1)}\chi^\rho_{1/2},\\[-2pt]
  \nonumber |D^{(\Xi)(2)}_{1/2}\rangle &=&\phi^{\rho(2)}\chi^\lambda_{1/2}
  -\phi^{\lambda(2)}\chi^\rho_{1/2},\\[-2pt]
  \nonumber  |D^{(\Xi)(1)}_{-1/2}\rangle&=&\phi^{\rho(1)}\chi^\lambda_{-1/2}
  -\phi^{\lambda(1)}\chi^\rho_{-1/2},\\[-2pt]
  |D^{(\Xi)(2)}_{-1/2}\rangle
  &=&\phi^{\rho(2)}\chi^\lambda_{-1/2}-\phi^{\lambda(2)}\chi^\rho_{-1/2},
  \label{eqB3}
\end{eqnarray}
and those of the quartet states
\begin{eqnarray}
  \nonumber |Q^{(\Xi)}_{3/2}\rangle &=&\phi^A\chi^S_{3/2},\\[-2pt]
  \nonumber |Q^{(\Xi)}_{1/2}\rangle &=&\phi^A\chi^S_{1/2},\\[-2pt]
  \nonumber |Q^{(\Xi)}_{-1/2}\rangle &=&\phi^A\chi^S_{-1/2},\\[-2pt]
  |Q^{(\Xi)}_{-3/2}\rangle &=&\phi^A\chi^S_{-3/2}.
\label{eqB4}
\end{eqnarray}
When two electrons with opposite spins share the same orbit, i.e., $N_1= N_2
\neq N_3 $ or $N_1\neq N_2=N_3$ or $N_1=N_3\neq N_2$,
\begin{eqnarray} 
  \nonumber |D^{(\Xi)(3)}_{1/2}\rangle &=&\phi^{\rho}\chi^\lambda_{1/2}
  -\phi^{\lambda}\chi^\rho_{1/2},\\[-2pt]
  |D^{(\Xi)(3)}_{-1/2}\rangle
  &=&\phi^{\rho}\chi^\lambda_{-1/2}-\phi^{\lambda}\chi^\rho_{-1/2}. 
\label{eqB5}
\end{eqnarray}
Here, \{$\phi$\} are the corresponding orbital wavefunctions. The superscript
``A'' on the right hand side of Eq.\,(\ref{eqB4}) describes the permutation
antisymmetric character for the exchange of any two
electrons and the ``$(1)$'' and ``$(2)$'' in the superscripts of
Eq.\,(\ref{eqB3}) are used to distinguish the
two doublet configurations with the same spin magnetic
quantum numbers or the two orbital functions with the same symmetry. The orbital
wavefunctions can be expressed as
\begin{widetext}
  \begin{eqnarray}\left\{
\begin{array}{llllll}
  \phi^{A}& =&\tfrac{1}{\sqrt{6}}(|N_1N_2N_3 \rangle-|N_1N_3N_2 \rangle+
  |N_2N_3N_1\rangle-|N_2N_1N_3 \rangle+|N_3N_1N_2 \rangle-|N_3N_2N_1\rangle)
  \\[2pt] 
  \phi^{\rho(1)} &=& \tfrac{1}{2\sqrt{2}}(|N_3N_2N_1\rangle-|N_2N_3N_1\rangle+
  |N_3N_1N_2\rangle-|N_1N_3N_2 \rangle)\\[2pt] 
  \phi^{\lambda(1)} &=&\tfrac{1}{2\sqrt{6}}(|N_1N_3N_2\rangle+|N_2N_3N_1\rangle+|N_3N_2N_1\rangle+|N_3N_1N_2\rangle-2|N_1N_2N_3 \rangle-2 |N_2N_1N_3\rangle)\\[2pt] 
  \phi^{\rho(2)} &=& \tfrac{1}{2\sqrt{6}}(2|N_1N_2N_3\rangle-2|N_2N_1N_3 \rangle-|N_2N_3N_1 \rangle+|N_3N_2N_1\rangle
  -|N_3N_1N_2\rangle+|N_1N_3N_2\rangle)\\[2pt] 
  \phi^{\lambda(2)} &=&
  \tfrac{1}{2\sqrt{2}}(|N_3N_1N_2\rangle-|N_3N_2N_1\rangle+|N_1N_3N_2\rangle-|N_2N_3N_1\rangle),
\end{array}\right.
\label{eqB6}
\end{eqnarray}
\begin{eqnarray}
  \phi^{\rho}= \left\{
    \begin{array}{llllll}
      \tfrac{1}{2}(|N_1N_3N_1 \rangle-|N_3N_1N_1 \rangle),&&N_1=N_2 \neq N_3,  \\[2pt]
      \tfrac{1}{2}(|N_1N_2N_2 \rangle-|N_2N_1N_2\rangle),&&N_1 \neq N_2=N_3, \\ [2pt]  
      \tfrac{1}{2}(|N_1N_2N_1\rangle-|N_2N_1N_1\rangle),&&N_1=N_3 \neq N_2 ,
    \end{array} \right.
\label{eqB7}
\end{eqnarray}
\begin{eqnarray}
   \phi^{\lambda} = \left\{
    \begin{array}{llllll}
      \tfrac{1}{2\sqrt{3}}(2|N_1N_1N_3\rangle-|N_1N_3N_1\rangle-|N_3N_1N_1\rangle),N_1=N_2\neq N_3, \\[2pt]  
      \tfrac{1}{2\sqrt{3}}(|N_2N_1N_2\rangle+|N_1N_2N_2\rangle-2|N_2N_2N_1\rangle), N_1 \neq N_2=N_3,\\[4pt]  
      \tfrac{1}{2\sqrt{3}}(2|N_1N_1N_2\rangle-|N_1N_2N_1 \rangle-|N_2N_1N_1\rangle),N_1=N_3 \neq N_2 .
    \end{array} \right.
\label{eqB8}
\end{eqnarray}

\section{MATRIX ELEMENTS OF COULOMB INTERACTION}\label{appC}
The Coulomb interaction components read\cite{Shen1,Lwang1,Lwang2}
\begin{equation} 
  \langle N_1N_2 N_3|H^{ij}_{\rm C}|N_1^\prime N_2^\prime N_3^\prime\rangle = 
  \frac{e^2}{16\pi^2\varepsilon_0\kappa}\delta_{N_k,N_k^\prime}\delta_{l_i+l_j,l_i'+l_j'}
\sum_{\gamma_i,\gamma_j,\gamma'_i,\gamma'_j=z,\overline{z}}
\eta^{\gamma_i}_{n_{{\rm v}i}}\eta^{\gamma_j}_{n_{{\rm
      v}j}}\eta^{\gamma'_i}_{n'_{{\rm v}i}}
\eta^{\gamma'_j}_{n'_{{\rm v}j}}
U(\phi^{\gamma_i}_{n_il_in_{zi}},\phi^{\gamma_j}_{n_jl_jn_{zj}},\phi^{\gamma'_i}_{n'_il'_in'_{zi}},
\phi^{\gamma'_j}_{n'_jl'_jn'_{zj}})
  \label{C1}
\end{equation}
with $\{i,j,k\}=\{1,2,3\}$.
Here, $|N_i \rangle \equiv |n_il_in_{zi}n_{{\rm v}i} \rangle$ is the
single-electron wavefunction of the $i$-th electron. Superscripts
$\gamma_{i(j)}$ and $\gamma'_{i(j)}$ run over the two valleys with 
$\eta^z_{\pm}=1$ and
$\eta^{\overline{z}}_{+}=-\eta^{\overline{z}}_{-}=1$ (Ref.\,\onlinecite{Lwang1}).
  $U$ in Eq.~(\ref{C1}) can be expressed as\cite{Lwang1,Shen1}
\begin{eqnarray} 
  \nonumber
  U(\phi^{\gamma_1}_{n_1l_1n_{z1}},\phi^{\gamma_2}_{n_2l_2n_{z2}},
  \phi^{\gamma'_1}_{n'_1l'_1n'_{z1}},\phi^{\gamma'_2}_{n'_2l'_2n'_{z2}})
  =\int^{\infty}_0\mathrm{d}k_{\parallel}\int^{\infty}_{-\infty}
  \mathrm{d}k_zk_{\parallel}P^{n'_1l'_1}_{n_1l_1}(k_{\parallel}) 
  P^{n_2l_2}_{n'_2l'_2}(k_{\parallel})\frac{W^{n'_{z_1}\gamma'_1}_{n_{z_1}\gamma_1}(k_z)
    [W^{n_{z_2}\gamma_2}_{n'_{z_2}\gamma'_2}(k_z)]^*}{k^2}.\\ 
  \label{eqC2}
\end{eqnarray}
$P^{n'l'}_{nl}$ can be obtained from the lateral part of the
integral $\langle nln_z\gamma|e^{i{\mathbf k}\cdot{\mathbf r}}|n'l'n'_z\gamma'\rangle$.
\begin{eqnarray} 
  P^{n'l'}_{nl}&=&\sqrt{\frac{n!n'!}{(n+|l|)!(n'+|l'|)!}}
  e^{-\frac{k^2_{\parallel}}{4\alpha^2}}\sum_{i=0}^{n'} 
  \sum_{j=0}^{n}C^i_{n',|l'|}C^j_{n,|l|}\overline{n}!L^{|l-l'|}_{\overline{n}}(\frac{k^2_{\parallel}}{4\alpha^2})[{\rm
    sgn}(l'-l)\frac{k_{\parallel}}{2\alpha}]^{|l'-l|}, 
 \label{eqC3}
\end{eqnarray}
with $C^i_{n,l}=\frac{(-1)^i}{i!}\binom{n+l}{n-i}$,
$\overline{n}=i+j+\frac{|l|+|l'|-|l-l'|}{2}$ and ${\rm sgn}(x)$ being the sign
function. 
 $W^{n'_{z}\gamma'}_{n_{z}\gamma}$ in Eq.\,(\ref{eqC2}) is the integral $\langle 
n_z\gamma|{\rm exp}(ik_zz)|n'_z \gamma'\rangle$.
\end{widetext}

\end{appendix}


\begin{thebibliography}{0}
\bibitem{square1} We calculate ground state energies of one, two and three
  electrons to obtain $\Delta
  \mu_i$ ($N_i=1,2$) via our model with $d_0$ being a fitting
  parameter. We choose $d_0$ which minimizes
  $\sqrt{(\Delta\mu_1-\Delta\mu^*_1)^2+(\Delta\mu_2-\Delta\mu^*_2)^2}$
  with $\Delta\mu^*_i$
  representing the experimental value.
\bibitem{width} In fact, from the fitting of the valley splitting, one can obtain
  many solutions of the well width because of the oscillation of the
  valley splitting against well width. However, the variation of the well width
  only induce slight effect on the addition energies, e.g., within $\sim 0.1$\,\%
  modification when the well width of Si4 varies $2$~nm. In our calculation, we
  choose one of these solutions near the suggested half-well width ($\sim
  4$~nm). For the same reason,
  we use the same well width in Si1 and Si2 as that in Si3 as a reasonable
  approximation.
\bibitem{orbit} From the explicit expression of the three-electron Hamiltonian,
  one can demonstrate that the Hamiltonian without spin-orbit term is non-vanishing only
  between states with the same total orbital angular momentum
  $L=l_1+l_2+l_3$. And the spin-orbit coupling is very weak compared
  with the orbital energy
  and Coulomb interaction. Therefore, one can denote each
  state (except in the vicinity of the anticrossing points) by the total orbital
  angular momentum.
\bibitem{nv} Here we take $n_{v1}=n_{v2}=-$ and $n_{v3}=+$
  for the permutation
  antisymmetric or symmetric characters of the orbital part of
  the basis functions for the electrons ``1'' and
  ``2''. Even though, the identity among the three
  electrons is still guaranteed  by the orthogonality and
  completeness of the entire Hilbert space.
\end{thebibliography}
\end{document}